\documentclass[journal]{IEEEtran}

\usepackage{amsmath, mathrsfs, mathtools}
\usepackage{amsfonts}
\usepackage{amssymb}
\usepackage{color}
\usepackage{multirow}
\usepackage{stmaryrd}
\usepackage{yfonts}
\usepackage{mathabx}
\usepackage{float}
\usepackage{stfloats}
\usepackage{amsfonts}
\usepackage{tcolorbox}
\usepackage[numbers,sort&compress]{natbib}
\usepackage{multicol}
\usepackage{algorithm}

\usepackage{algorithmic}

\usepackage{booktabs}

\usepackage{pifont}
\usepackage{varwidth}
\usepackage{afterpage}
\usepackage{balance}
\usepackage{lipsum}

\usepackage{tikz, pgfplots,graphicx,xcolor}
\usetikzlibrary{plotmarks,spy,backgrounds}
\pgfplotsset{compat=newest}

\usepackage{mathbbol}

\def\mindex#1{\index{#1}}

\def\sq{\hbox{\rlap{$\sqcap$}$\sqcup$}}
\def\qed{\ifmmode\sq\else{\unskip\nobreak\hfil
\penalty50\hskip1em\null\nobreak\hfil\sq
\parfillskip=0pt\finalhyphendemerits=0\endgraf}\fi\medskip}

\long\def\defbox#1{\framebox[.9\hsize][c]{\parbox{.85\hsize}{%
\parindent=0pt
\baselineskip=12pt plus .1pt      
\parskip=6pt plus 1.5pt minus 1pt 
 #1}}}

\long\def\beginbox#1\endbox{\subsection*{}%
\hbox{\hspace{.05\hsize}\defbox{\medskip#1\bigskip}}%
\subsection*{}}

\def\endbox{}

\def\diag{{\text{diag}}}

\def\tr{\mathsf{tr}}

\newsavebox{\junk}
\savebox{\junk}[1.6mm]{\hbox{$|\!|\!|$}}

\def\det{{\mathop{\rm det}}}

\def\argmin{\mathop{\rm arg\, min}}

\def\argmax{\mathop{\rm arg\, max}}

\def\Re{\field{R}}

\def\bB{{\mathbb B}}
\def\bC{{\mathbb C}}

\def\bE{{\mathbb E}}

\def\bI{{\mathbb I}}

\def\bR{{\mathbb R}}
\def\bS{{\mathbb S}}

\def\bfmath#1{{\mathchoice{\mbox{\boldmath$#1$}}%
{\mbox{\boldmath$#1$}}%
{\mbox{\boldmath$\scriptstyle#1$}}%
{\mbox{\boldmath$\scriptscriptstyle#1$}}}}

\def\bfmY{\bfmath{Y}}

\def\bfmhhaY{\bfmath{\hhaY}} 
\def\bfmhhaY{\hbox to 0pt{$\widehat{\bfmY}$\hss}\widehat{\phantom{\raise 1.25pt\hbox{$\bfmY$}}}}

\def\til={{\widetilde =}}

 \def\FRAC#1#2#3{\genfrac{}{}{}{#1}{#2}{#3}}

\def\ddtp{{\mathchoice{\FRAC{1}{d^{\hbox to 2pt{\rm\tiny +\hss}}}{dt}}%
{\FRAC{1}{d^{\hbox to 2pt{\rm\tiny +\hss}}}{dt}}%
{\FRAC{3}{d^{\hbox to 2pt{\rm\tiny +\hss}}}{dt}}%
{\FRAC{3}{d^{\hbox to 2pt{\rm\tiny +\hss}}}{dt}}}}

\def\average#1,#2,{{1\over #2} \sum_{#1}^{#2}}

\def\eye(#1){{\bf(#1)}\quad}

\newtheorem{remark}{{\bf Remark}}

\def\eq#1/{(\ref{e:#1})}

\newcommand{\beqn}[1]{\notes{#1}%
\begin{eqnarray} \elabel{#1}}

\newcommand{\eeqn}{\end{eqnarray} }

\newcommand{\beq}[1]{\notes{#1}%
\begin{equation}\elabel{#1}}

\newcommand{\eeq}{\end{equation}}

\def\bdes{\begin{description}}
\def\edes{\end{description}}

\newcounter{rmnum}

\newcounter{anum}

{\end{list}}

\def\ass(#1:#2){(#1\ref{#1:#2})}

\def\ritem#1{
\item[{\sf \ass(\current_model:#1)}]
}

\newenvironment{recall-ass}[1]{%
\begin{description}
\def\current_model{#1}}{
\end{description}
}

\long\def\comment#1{}

\newfont{\bb}{msbm10 scaled 1100}

\newcommand{\av}{{\bf a}}
\newcommand{\bv}{{\bf b}}
\newcommand{\cv}{{\bf c}}
\newcommand{\dv}{{\bf d}}
\newcommand{\ev}{{\bf e}}
\newcommand{\fv}{{\bf f}}
\newcommand{\gv}{{\bf g}}
\newcommand{\hv}{{\bf h}}

\newcommand{\nv}{{\bf n}}

\newcommand{\pv}{{\bf p}}

\newcommand{\rv}{{\bf r}}

\newcommand{\wv}{{\bf w}}
\newcommand{\vv}{{\bf v}}
\newcommand{\xv}{{\bf x}}
\newcommand{\yv}{{\bf y}}
\newcommand{\zv}{{\bf z}}

\newcommand{\Cm}{{\bf C}}

\newcommand{\Em}{{\bf E}}
\newcommand{\Fm}{{\bf F}}

\newcommand{\Hm}{{\bf H}}
\newcommand{\Id}{{\bf I}}
\newcommand{\Jm}{{\bf J}}

\newcommand{\Wm}{{\bf W}}

\newcommand{\alphav}{\hbox{\boldmath$\alpha$}}

\newcommand{\gammav}{\hbox{\boldmath$\gamma$}}

\newcommand{\lambdav}{\hbox{\boldmath$\lambda$}}

\newcommand{\muv}{\hbox{\boldmath$\mu$}}

\newcommand{\phiv}{\hbox{\boldmath$\phi$}}

\newcommand{\thetav}{\hbox{\boldmath$\theta$}}

\newcommand{\xiv}{\hbox{\boldmath$\xi$}}
\newcommand{\sigmav}{\hbox{\boldmath$\sigma$}}

\newcommand{\Gammam}{\hbox{\boldmath$\Gamma$}}
\newcommand{\Lambdam}{\hbox{\boldmath$\Lambda$}}

\newcommand{\Sigmam}{\hbox{\boldmath$\Sigma$}}
\newcommand{\Phim}{\hbox{\boldmath$\Phi$}}

\renewcommand{\det}{{\hbox{det}}}
\newcommand{\trace}{{\hbox{tr}}}
\renewcommand{\arg}{{\hbox{arg}}}

\renewcommand{\Re}{{\rm Re}}
\renewcommand{\Im}{{\rm Im}}

\newcommand{\herm}{{\sf H}}

\newcommand{\transp}{{\sf T}}
\renewcommand{\vec}{{\rm vec}}

\allowdisplaybreaks
\newcommand{\normd}[1]{{\left\vert\kern-0.25ex\left\vert\kern-0.25ex\left\vert #1 
		\right\vert\kern-0.25ex\right\vert\kern-0.25ex\right\vert}}

\title{ Low-Complexity  Near-Field  Channel Estimation and Subcarrier-Cooperative Hybrid Precoding for Wideband XL-MIMO OFDM Systems }

\author{Kangda Zhi, Tianyu Yang,   and Giuseppe Caire, {\itshape Fellow, IEEE \upshape}		
		\thanks{  Kangda Zhi is with Guangdong Provincial Key Laboratory of Aerospace Communication and Networking Technology, Harbin Institute of Technology (Shenzhen), Shenzhen, China, 518055 (email: zhikangda@hit.edu.cn).}
	\thanks{Tianyu Yang and Giuseppe Caire are with Communications and Information Theory Group (CommIT), Technische Universit\"{a}t Berlin, 10587 Berlin, Germany (e-mail:  caire@tu-berlin.de).}

 \vspace{-20pt}
}

\begin{document}
\maketitle

\begin{abstract}
This paper addresses both accurate near-field channel acquisition and scalable precoding for wideband extremely large aperture MIMO (XL-MIMO) OFDM systems, with particular focus on reducing complexity. In the considered system, each MIMO multipath component is characterized by five continuous angle-distance-delay parameters, making conventional sparse recovery methods computationally infeasible while suffering from grid mismatch. We propose a decoupled off-grid channel estimation algorithm based on sequential sparse Bayesian learning (SBL). By exploiting the separable structure of the OFDM near-field atom, the five-dimensional joint search is replaced by a sequence of low-dimensional operations and active-set inference, avoiding multiplicative scaling with the per-dimension grid sizes. A continuous-domain Newton refinement is incorporated based on the exact marginal likelihood to mitigate the grid mismatch.  Based on the estimated multipath parameters, we then develop a subcarrier-cooperative hybrid precoding framework for multiuser wideband transmission. A large-scale-matrix-inversion-free alternating optimization algorithm is established to maximize the sum user rate, together with a non-iterative low-complexity design that exploits the parameterized channel structure and provides an effective initialization. Simulation results demonstrate the accuracy of the estimation methods compared to several benchmarks and the effectiveness of the precoding algorithm based on estimated channel parameters.
\end{abstract}


\section{Introduction}
Research towards 6G has put emphasis on extra-large-scale massive MIMO (XL-MIMO) with mid-band spectrum, spanning sub-6 GHz and frequency range 3 (FR3). Benefiting from the large array aperture and broader spectral resources, wideband XL-MIMO can exploit unprecedented spatial resolution and beamforming (BF) gains to achieve high communication capacity and sensing accuracy. Developing channel estimation and BF algorithms for wideband XL-MIMO with legacy orthogonal frequency division multiplexing (OFDM) is therefore of significant importance.

Evolving from narrowband MIMO and conventional MIMO-OFDM systems, XL-MIMO OFDM introduces two fundamental challenges to algorithm design. The first is how to acquire accurate near-field wideband channel state information (CSI) \cite{long2026channel}. Owing to the large array aperture, signal propagation from XL-MIMO enters the near-field regime, where the conventional planar-wave assumption breaks down and spherical wavefronts should be considered \cite{Lu2024Survey}. As a result, conventional DFT codebooks suffer from model mismatch, where the energy of each near-field path will spread across multiple angular bins. Moreover, the wideband frequency-selective propagation further introduces delay parameters, expanding the channel representation to the angle-distance-delay domain. Therefore, near-field wideband channel estimation faces the trouble of dramatically increased complexity and overhead associated with the high-dimensional grids of channel propagation parameters.

To tackle the channel estimation problem, angle-distance domain dictionaries/codebooks have been developed for multiple-input single-output (MISO) systems  \cite{cui2022channel}. For wideband OFDM, the authors of \cite{wang2026compressive} established a frequency-independent orthogonal dictionary to recover the wideband MISO channel. A distance–frequency-invariant codebook was proposed in \cite{yuan2026low} for uniform circular arrays with experimental validation. Besides, the authors in \cite{chen2026near} proposed a total variation-regularized block sparse Bayesian learning (SBL) method to estimate the base station (BS)-side angle-distance parameters for a uniform linear array (ULA)-equipped BS. Meanwhile, data-driven methods, such as denoising diffusion models \cite{feng2026near} and dual-attention-aided deep-unfolded SBL \cite{li2026wideband}, were exploited to reconstruct the wideband channel matrices with ULA and uniform planar arrays (UPA), respectively.

Beyond MISO, the estimation of wideband MIMO channels in the near field is more challenging, since multi-antenna users further exacerbate the searching dimension of channel parameters. This problem has been considered in \cite{thallapalli2026gridless}, which, however, suffers from prohibitively high complexity. Some reconfigurable intelligent surface-related near-field wideband MIMO channel estimation has been studied \cite{tuo2026near,yang2024near}, which relies on several approximations and has grid mismatch. Motivated by these limitations, our objective is to enable low-complexity while accurate off-grid estimation for near-field MIMO-OFDM channels. We leverage the sequential SBL framework \cite{tipping2003fast,pote2025theory} 
that maintains a very small active set with inference performed in a low-dimensional subspace.  
By further exploiting the separable factored form of the XL-MIMO-OFDM channel, we propose an efficient coarse-grid search followed by an off-grid refinement scheme, 
avoiding the complexity incurred by processing the full five-dimensional angle-distance-delay dictionary.


The second challenge of XL-MIMO-OFDM is how to realize high-capacity yet computationally efficient multi-user wideband BF. In fact, this issue is strongly coupled with the stage of channel estimation, since the precoding should be customized based on the acquired information of channels (i.e., the acquisition of the whole channel matrix or multipath parameters). XL-MIMO BF suffers from an extremely large number of design variables together with multiple users and subcarriers. This poses a key requirement to restrict computational complexity and construct reliable initial solutions.

For point-to-point wideband XL-MIMO, a structured OFDM modulation scheme with spatial- and frequency-wideband effects has been proposed in \cite{huang2024structured}. Holographic wideband precoding with multiple users has been studied in \cite{cheng2026multi}. Besides, works \cite{wang2024beamfocusing,ding2025near} have investigated beamfocusing design under the beam-splitting effect with true time delay. However, this effect is not severe for the mid-band spectrum. Near-field wideband precoding under spatial non-stationarity was studied as well \cite{liu2024hybrid}. Nevertheless, the precoding design in these contributions is not directly supported by the channel estimation results. In this work, we develop a wideband hybrid BF exactly enabled by the estimated wideband MIMO-OFDM channels and with low complexity. Particularly, we investigate an interesting scheme called subcarrier-cooperative transmission \cite{hellings2011inseparability} that jointly processes signals across subcarriers, thereby offering additional cross-subcarrier degree of freedom (DoF) to suppress interference. Therefore, our scheme has the potential to fully unleash the capabilities of wideband XL-MIMO to support the extremely high data rates envisioned for 6G applications.

The main contributions of our low-complexity XL-MIMO-OFDM estimation-and-transmission framework are summarized as follows.
	\begin{itemize}
    \item We investigate a very challenging scenario where a UPA XL-MIMO communicates with multi-antenna users in OFDM wideband systems. The first task is to accurately estimate the five-dimensional angle-distance-delay parameterized near-field channel with low complexity. The second task is to realize high-rate communication based on the available CSI with low precoding complexity, given a large number of antennas, multiple users, and multiple subcarriers. 
	\item For near-field wideband channel estimation, we propose a decoupled off-grid sequential SBL method. We exploit the separable structure of near-field OFDM atoms to decompose high-dimensional searches into a sequence of low-dimensional operations, avoiding the multiplicative complexity with respect to five-dimensional grids. A continuous-domain Newton refinement based on marginal likelihood is further incorporated to eliminate grid mismatch and improve resolution. Notably, we show that all posterior updates of our sequential SBL-based method can be realized using Schur-complement, completely eliminating explicit matrix inversions. Beyond the considered channel-estimation application, we emphasize that our method offers a  scalable sequential Bayesian framework for multidimensional continuous sparse estimation with separable atom structures.
	\item For wideband multi-user downlink (DL) data transmission, we propose a subcarrier-cooperative transmission scheme based on the estimated channel parameters, which has additional cross-subcarrier DoF for interference suppression. To maximize the sum user rate, a large-scale-matrix-inversion-free alternating optimization framework is first established for subcarrier-cooperative hybrid precoding. Then, a low-complexity sub-optimal scheme is proposed without alternating iterations based on estimated multipath parameters.
	\item Simulations demonstrate the accuracy of the channel estimation algorithms compared to benchmarks and Cram\'er--Rao lower bound (CRLB). The effectiveness and benefits of the proposed subcarrier-cooperative hybrid precoding are also validated under the estimated CSI.
    \end{itemize}


\section{System Model}
\begin{figure}
    \centering
    \includegraphics[width=0.8\linewidth]{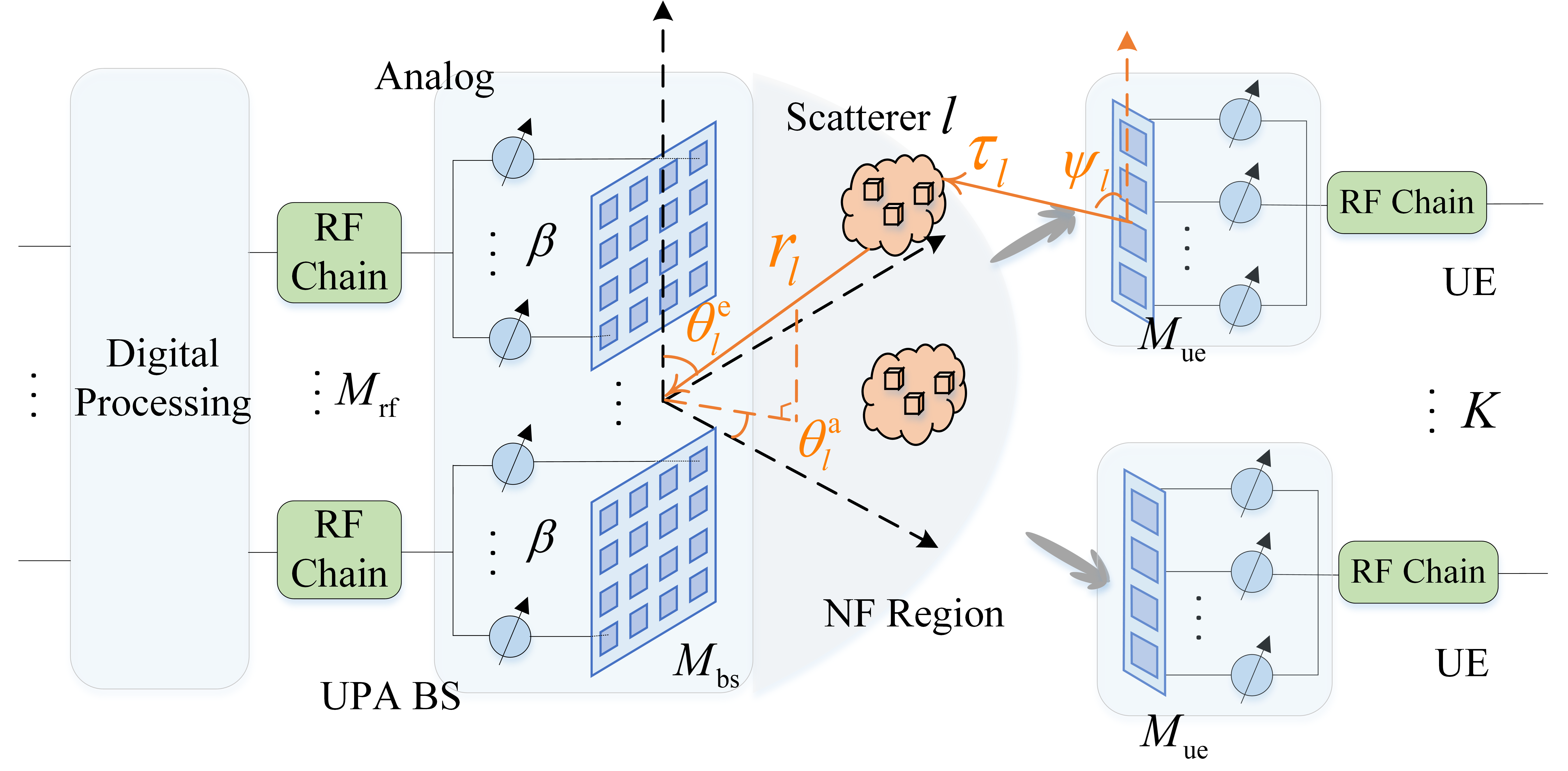}
         \vspace{-3mm}
    \caption{Illustration of the considered XL-MIMO OFDM near-field system.}
    \label{fig:system_model}
\end{figure}

As shown in Fig.~\ref{fig:system_model}, we consider an XL-MIMO OFDM communication system, where the BS employs a UPA of $M_{\rm bs} = M_{\rm h} M_{\rm v}$ ($M_{\rm h}$ horizontal and $M_{\rm v}$ vertical) antennas, and a generic user equipment (UE) is equipped with $M_{\rm ue}$ antennas in ULA. Due to the hardware complexity, we consider a single RF chain at the UE and a sub-connected hybrid architecture at the BS, where $M_{\rm bs}$ antenna elements connect to $M_{\rm rf}$ radio-frequency (RF) chains, with $M_{\rm rf} \ll M_{\rm bs}$ and $\beta \triangleq M_{\rm bs} / M_{\rm rf}$ being the number of antennas connected to each RF chain.
The bandwidth of the wideband system is divided into $N$ subcarriers with subcarrier spacing $\Delta f$.

\subsection{Channel Model}
We focus on the systems with sub-6 GHz or low frequency range of FR3, where the cell size is still relatively large and the LoS 
probability is therefore low. Hence, we consider a multipath channel with $L$ NLoS paths. Additionally, due to the large antenna aperture at the BS, scatterers fall inside the Fresnel region of the BS array, i.e., we need to consider the near-field channels between the BS and the scatterers. From the UE side, the scatterers are assumed to be located at far field due to the small aperture of the UE array.  Then, the channel at subcarrier $n$ is given as 
\begin{align}\label{eq:channel}
    \Hm[n] = \sum^L_{l=1} \alpha_l  g_n(\tau_l) \av(\theta^{\rm a}_l, \theta^{\rm e}_l, r_l) \bv^\herm(\psi_l) \in \bC^{M_{\rm bs}\times M_{\rm ue}},
\end{align}
where $\alpha_l, \tau_l, \psi_l, (\theta^{\rm a}_l, \theta^{\rm e}_l,r_l)$ denote the complex gain, propagation delay, UE angle, and BS-side parameters (azimuth angle, elevation angle, distance) of the $l$-th path, respectively. $g_n(\tau_l) = \exp(-j 2 \pi (n-1) \Delta f \tau_l)$ is the phase shift due to path delay $\tau_l$ at subcarrier $n$. $\bv(\psi_l) \in \bC^{M_{\rm ue}}$ is the ULA steering vector whose $m$-th element is $[\bv(\psi_l)]_m = \exp(-j\pi (m-1) \sin (\psi_l))$ under half-wavelength antenna spacing. $\av(\theta^{\rm a}_l, \theta^{\rm e}_l, r_l)\in \bC^{M_{\rm bs}}$ is the near-field spherical-wave steering vector with entries
\begin{align}
    [\av(\theta^{\rm a}_l, \theta^{\rm e}_l, r_l)]_m = \exp\left(-jk_0 \|\pv_m^{\rm bs}-\pv^{\rm sc}_l\|\right),
\end{align}
where $k_0=\frac{2\pi}{\lambda_c}$ with $\lambda_c$ being the carrier wavelength,\footnote{We do not consider the beam squint effect due to low ratio between bandwidth and carrier frequency in the considered frequency band.} $\pv_m^{\rm bs}$ and $\pv_l^{\rm sc}$ are locations of $m$-th antenna of the BS and the $l$-th scatterer, respectively.

\subsection{Transmit and Received Pilot Signal}
We assume that the UE transmits pilot signal over $T$ slots. The pilot signal at time $t$ is given as $\sqrt{P_{\rm ue}} \fv_t$, where $P_{\rm ue}$ is the UE transmit power, and $\fv_t\in \bC^{M_{\rm ue}}$ is the analog precoder with constant modulus constraints $|[\fv_t]_i|=1/\sqrt{M_{\rm ue}}, i\in[M_{\rm ue}]$.\footnote{In this paper, for a positive integer $N$, we define $[N] \triangleq \{1,\dots, N\}$.} The BS applies an analog combiner $\Wm_t \in \bC^{M_{\rm bs} \times M_{\rm rf}}$ at time $t$. Due to the sub-connected structure, the analog combiner is a block diagonal matrix, i.e., $\Wm_t = {\sf blkdiag} (\wv_{1,t},\dots, \wv_{M_{\rm rf},t})$, where $\wv_{m,t} \in \bC^{\beta}$ is the analog combining vector of the $m$-th RF chain at time $t$, with constant modulus, i.e., $|[\wv_{m,t}]_i| = 1/\sqrt{\beta}, i\in[\beta]$, and $\Wm_t^\herm \Wm_t = \Id_{M_{\rm rf}}$. 
Then, the received pilot signal by the BS at time $t$ and subcarrier $n$ is given as
\begin{align}\label{eq:received_y}
    \yv_t[n] = \sqrt{P}\Wm_t^\herm \Hm[n]\fv_t + \Wm_t^\herm \zv_t[n] \in \bC^{M_{\rm rf}},
\end{align}
where $\zv_t \sim\mathcal{CN}(\mathbf{0},\sigma^2 \Id_{M_{\rm bs}})$ is the additive white Gaussian noise (AWGN) with uplink (UL) noise power $\sigma^2$.

\subsection{Challenge of Full Channel Estimation}
Given pilot observations and known pilots $\{\yv_t[n], \Wm_t, \fv_t \mid t\in [T],n\in[N]\}$ at the BS, the task is to estimate the full wideband channel vector $\hv = \vec(\mathcal{H})\in \bC^{\widetilde{M}}$, where $\mathcal{H}\in \bC^{M_{\rm bs}\times M_{\rm ue}\times N}$ is the three-dimensional (3-D) array of $\{\Hm[n] \mid n\in[N]\}$ and $\widetilde{M}\triangleq M_{\rm bs}M_{\rm ue}N$. Although the full wideband channel has an extremely large dimension $\widetilde{M}$, it is fully determined by much fewer parameters of the multipath, making it logical to estimate the multipath parameters. Specifically, collecting all observations into a 3-D array $\mathcal{Y}\in \bC^{M_{\rm rf}\times N\times T}$ and vectorizing it as  $\yv = \vec(\mathcal{Y})\in\bC^{M}$ with $M \triangleq M_{\rm rf}NT$, 
the received signal can be rewritten as 
\begin{align}
    \yv &= \sum^L_{l=1} \alpha_l \phiv(\xiv_l) + \widetilde{\zv} =\widetilde{\Phim} \alphav + \widetilde{\zv} \in \bC^M, \label{eq:linear_form} 
\end{align}
where $\xiv_l \triangleq [\tau_l, \psi_l, \theta^{\rm a}_l, \theta^{\rm e}_l, r_l]^\transp\in \bR^5$ contains 5-D continuous parameters of each path, $\phiv(\xiv_l)\in\bC^M$ is the atom corresponding to parameter $\xiv_l$, $\widetilde{\Phim} \triangleq [\phiv(\xiv_1),\dots, \phiv(\xiv_L)]$ is the matrix of all $L$ atoms, $\alphav\triangleq[\alpha_1,\dots,\alpha_L]$, and $\widetilde{\zv} \sim \mathcal{CN}(\mathbf{0}, N_0\Id_M)$ is the corresponding effective UL noise with $N_0 = \sigma^2/P_{\rm ue}$.  

Given the standard linear form in \eqref{eq:linear_form}, a common method is to discretize the parameter space into a finite set of grid points and apply compressed sensing methods, such as Orthogonal Matching Pursuit (OMP) and SBL. 
However, direct application of such methods requires a joint dictionary over a 5-D continuous parameter space $(\tau, \psi,\theta^{\rm a},\theta^{\rm e},r)$, leading to unaffordable computational complexity and severe memory overflow. 
Specifically, by discretizing the parameter space into $G=G_\tau G_\psi G_{\theta^{\rm a}} G_{\theta^{\rm e}} G_r$ joint grid points with $G_\tau, G_\psi, G_{\theta^{\rm a}}, G_{\theta^{\rm e}}, G_r$ being per-dimension grid sizes, we construct a dictionary matrix $\Phim\triangleq[\phiv_1, \dots, \phiv_G]\in \bC^{M\times G}$, where $\phiv_i\triangleq \phiv(\xiv^{\rm g}_i), i\in[G]$ is the $i$-th atom with $\xiv^{\rm g}_i$ being the atom position of the $i$-th grid point including 5-D parameters.  Then, the received signal in \eqref{eq:linear_form} can be approximated as  
\begin{align}
        \yv \approx \Phim \xv + \widetilde{\zv},
\end{align}
where $\xv\in \bC^{G}$ is the unknown  vector of $G$ grid points.

\section{On-Grid Sequential SBL Framework}\label{sec:on-grid-SBL}
To circumvent the curse of dimensionality and keep good estimation performance, we adopt the sequential SBL approach that avoids computationally intensive global operations over all atoms in the classical Expectation–Maximization (EM)-based SBL algorithm \cite{wipf2004sparse}. Sequential SBL searches for the single most informative atom, where we will later show that the search can be decomposed into a sequence of cheap low-dimensional operations by exploiting the separable structure of the OFDM MIMO channels.

\subsection{Sequential SBL}
SBL treats elements in $\xv$ as random independent variables that follow a parameterized Gaussian prior with mean zero and uncorrelated components, i.e., 
\begin{align}
    \xv \sim \mathcal{CN}(\mathbf{0}, \Gammam),
\end{align}
where $\Gammam = \diag(\gammav) = \diag([\gamma_1, \dots, \gamma_G])$ with $\gamma_i \geq 0$ being the hyperparameter of $i$-th grid point. 
SBL follows the Type-II estimation framework that solves the problem in the hyperparameter $\gammav$-space by minimizing the minus marginal log-likelihood function\footnote{In this work, we assume that the noise power $\sigma^2$ is known at the BS. However, the proposed method can be easily extended to jointly estimate the channel parameters and noise power.} 
\begin{align}\label{eq:sbl-cost}
    \min_{\gammav} \; \mathcal{L}(\gammav) \triangleq -\log p(\yv|\gammav)= \log \det (\Cm)  + \yv^\herm \Cm^{-1} \yv,
\end{align}
where $\Cm = \Phim\Gammam\Phim^\herm + N_0\Id_M$ is the covariance matrix of $\yv$. Once $\gammav$ is estimated by solving $\eqref{eq:sbl-cost}$, the target variable $\xv$ is estimated via the posterior mean of the posterior parameter distribution conditioned on the observation $p(\xv|\yv,\gammav)$, which can be obtained by combining the likelihood and prior within the rule of Bayes. The resulting posterior distribution is $p(\xv|\yv,\gammav)\sim\mathcal{CN}(\muv,\Sigmam)$, where the posterior covariance  and mean are respectively given as \cite{tipping2003fast}
\begin{align}\label{eq:posterior_cov_mean}
    \Sigmam = (N_0^{-1} \Phim^\herm\Phim + \Gammam^{-1})^{-1},\quad
    \muv =N_0^{-1}\Sigmam \Phim^\herm \yv.
\end{align}

The cost function in problem \eqref{eq:sbl-cost} is not convex and can be solved iteratively.  
One common approach is to use the EM method that requires a heavy matrix inversion with the dimension of the grid size in each iteration, resulting in unaffordable complexity under a large grid size. Unlike SBL-EM, sequential SBL iteratively adds a single grid point into the active set of grid points that maximally minimizes the cost function in \eqref{eq:sbl-cost}, and hence only requires a light matrix inversion with the small dimension of the active set size. Concretely, given an active index set $\bS\subset[G]$ with cardinality $S\triangleq |\bS|<G$,  we define $\gammav_{\bS}\in\bR^{S}$,  $\Phim_\bS\in \bC^{M\times S}$ and $\Cm_{\bS} \triangleq \Phim_\bS (\Gammam_\bS \triangleq \diag(\gammav_\bS))\Phim_\bS^\herm + N_0\Id_M$ as respectively the corresponding hyperparameter vector $\gammav$, sensing matrix $\Phim$ and covariance matrix $\Cm$ restricted to only the grid points of the active set $i\in\bS$.

Now, we evaluate the contribution of adding a grid point that is not in the active set $i\in [G] \setminus \bS$. The idea is to choose the new grid point that provides the maximum improvement of the evidence. Assuming that $\{\gamma_i|i\notin\bS\}$ are all initialized as zero, it is shown in \cite{pote2025theory} and \cite[eq. (18)]{tipping2003fast} that the change to the cost function {\eqref{eq:sbl-cost}} by adding a new grid point $i\notin \bS$ is given as
\begin{align}
    \Delta\mathcal{L}(\gamma_i) &\triangleq \mathcal{L}(\gammav_{\bS \cup\{i\}}) - \mathcal{L}(\gammav_{\bS}) \\
    &= \log(1+\gamma_i s_i) - \frac{\gamma_i |q_i|^2}{1+\gamma_i s_i}, \label{eq:delta_L}
\end{align}
where $s_i \triangleq \phiv_i^\herm \Cm_{\bS}^{-1}\phiv_i \geq 0$ and $q_i \triangleq \phiv_i^\herm \Cm_{\bS}^{-1}\yv \in \bC$ are so called \textit{sparsity factor} and \textit{quality factor}, respectively.
To choose a new grid point, we need to find $i\notin\bS$ and corresponding $\gamma_i$ that minimize the change $\Delta\mathcal{L}(\gamma_i)$, i.e.,\footnote{Note that our cost function is negative log-likelihood and thus we want to minimize the negative change. }
\begin{align}\label{eq:change_opt}
    i = \argmin_{i\notin\bS}\;\min_{\gamma_i\geq 0} \;\Delta\mathcal{L}(\gamma_i).
\end{align}
The optimal $\gamma_i$ is obtained by setting the derivative to zero, i.e., let $\Delta\mathcal{L}'(\gamma_i)=0$, which gives
\begin{align}\label{eq:gamma_opt}
    \gamma_i^\star = \max\left(\frac{|q_i|^2-s_i}{s_i^2},0\right).
\end{align}
Putting \eqref{eq:gamma_opt} back to \eqref{eq:delta_L}, we have non-zero change  
\begin{align}\label{eq:function_Q}
    \Delta\mathcal{L}(\gamma_i^\star) = \log(Q_i) - Q_i + 1, \; \text{when}\; Q_i>1,
\end{align}
where $Q_i\triangleq \frac{|q_i|^2}{s_i}$. Since we want to minimize \eqref{eq:function_Q} and  \eqref{eq:function_Q} is negative and strictly decreasing in $Q_i \geq 1$, the optimal grid point  maximizes $Q_i$, i.e.,
\begin{align}\label{eq:search_i}
    i^\star = \argmax_{i\notin\bS} \; Q_i.
\end{align}
If the optimal $Q_i$ is  $Q_{\hat{i}}\leq 1$, we have $\gamma^\star_{i^\star}=0$, indicating that no more new grid point should be added into the active set.  

\subsection{Analysis on Sequential SBL for Our Estimation Problem}\label{sec:analysis_sbl}
The dominant computational cost of sequential SBL is to solve \eqref{eq:search_i} in each iteration, which requires evaluating $q_i$ and $s_i$ for all candidates. Under a very large grid size, this evaluation can be very costly.
To tackle this, we first rewrite $q_i$ and $s_i$ in a computationally efficient form. Using the matrix inversion lemma to $\Cm_{\bS}^{-1}$, we have 
\begin{align}
    \Cm_{\bS}^{-1} &= N_0^{-1}\Id - N_0^{-2} \Phim_\bS (N_0^{-1}\Phim^\herm_\bS\Phim_\bS + \Gammam_\bS^{-1})^{-1}\Phim^\herm_\bS,\\
    &= N_0^{-1}\Id - N_0^{-2} \Phim_\bS \Sigmam_\bS\Phim^\herm_\bS. \label{eq:woodbury_C}
\end{align}
Using \eqref{eq:woodbury_C}, we can calculate $q_i$ and $s_i$ as 
\begin{align}
    q_i &= N_0^{-1}\phiv_i^\herm \yv - N_0^{-2}\phiv_i^\herm \Phim_\bS \Sigmam_\bS \Phim_\bS^\herm \yv, \\
    &= N_0^{-1}\phiv^\herm_i (\yv - \Phim_\bS \muv_\bS), \label{eq:q_mu} \\
    &= N_0^{-1}\phiv^\herm_i \rv_\bS, \label{eq:q_i} \\
    s_i &= N_0^{-1}\|\phiv_i\|^2 - N_0^{-2}\phiv_i^\herm \Phim_\bS \Sigmam_\bS \Phim_\bS^\herm  \phiv_i,\label{eq:s_i}
\end{align}
where we define $\Sigmam_\bS \triangleq (N_0^{-1}\Phim^\herm_\bS\Phim_\bS + \Gammam_\bS^{-1})^{-1}$ and $\muv_\bS\triangleq N_0^{-1}\Sigmam_\bS\Phim_\bS^\herm \yv$ as respectively the posterior covariance $\Sigmam$ and mean $\muv$ in \eqref{eq:posterior_cov_mean} with the contribution of grid points $i\notin\bS$ removed, and $\rv_\bS\triangleq \yv - \Phim_\bS \muv_\bS$ as the Bayesian residual based on active grid points in $\bS$. 

Now, we analyze the roles of the quality and sparsity factors that will guide our algorithm design. 
\begin{itemize}
    \item {\bf Analysis on $q(\xiv)$:} From \eqref{eq:q_i} we see that the quality factor is the inner product of the candidate atom $\phiv(\xiv)$ with the Bayesian residual $\rv_\bS$, whose modulus measures the orthogonality between the candidate atom and the residual. Thus, the candidate with higher $|q(\xiv)|^2$ aligns with the residual better.  
    \item {\bf Analysis on $s(\xiv)$:} From \eqref{eq:s_i} we see that the sparsity factor has two terms. The first term is the total energy of the candidate atom, which has the same
expectation across candidates under random constant-modulus pilots $\{\Wm_t, \fv_t\}$. The second term is the projected energy onto the subspace of the active set, weighted by $\Sigmam_\bS$. Therefore, $s(\xiv)$ measures the effective energy of a candidate, which plays as a penalty to avoid selecting candidates that have high correlation with the current active set. 
\end{itemize}
Furthermore, from a computational perspective, the atom $\phiv_i$ possesses a separable structure that decomposes the inner product into a sequence of low-dimensional operations, enabling low-cost computation of $q_i$ (see Section~\ref{sec:on-grid}). However, processing $s_i$ would significantly increase the searching cost due to the quadratic form in \eqref{eq:s_i}. Based on the aforementioned analysis on $s_i$, we propose to use a coarse grid with low mutual correlation of different grid points to avoid calculating $s_i$, and search for a new grid point only based on $q_i$. To mitigate the mismatch error due to the coarse grid and enhance the estimation performance, we then apply a local off-grid Newton refinement. At this off-grid stage, only a single candidate needs to be evaluated, so we can afford to use both $q(\xiv)$ and $s(\xiv)$, making the final acceptance be based on the exact evidence criterion, and guaranteeing that each addition strictly improves the marginal likelihood. As a result, the proposed algorithm preserves the monotonic convergence guarantee of sequential SBL.

\section{Decoupled Off-Grid Sequential SBL-Based Channel Estimation}
In this section, we describe the proposed decoupled off-grid sequential SBL-based channel estimation scheme. 
In each main iteration, we apply the on-grid search over a fixed coarse grid based on only the quality factor $q_i$. The chosen candidate is refined in the continuous domain by Newton steps and added to the active atom set if it improves the marginal likelihood. Since a specific grid point of a coarse grid may serve as the initial point of Newton refinement for multiple paths, each on-grid search is always performed on the same complete grid. Accordingly, we redefine the active atom set as $\bar{\bS} = \{(\xiv^{\rm act}_i\, \gamma_i^{\rm act})\}_{i=1}^{\bar{S}}$ that stores the positions and hyperparameters of $\bar{S}$ refined off-grid atoms, instead of grid point indices in the fully on-grid framework described in Section~\ref{sec:on-grid-SBL}.\footnote{Note that we use notation $i$ to index both the active atoms in $\bar{\bS}$ and the grid points in $[G]$.} Then, the corresponding values based on the active set, e.g., $\Cm_{\bar{\bS}}, \Phim_{\bar{\bS}}, \rv_{\bar{\bS}}$, are calculated with the contribution of atoms only in $\bar{\bS}$.

\subsection{Decoupled Coarse On-Grid Search}\label{sec:on-grid}
As explained in Section~\ref{sec:analysis_sbl}, we use $|q_i|^2$ instead of $Q_i$ in \eqref{eq:search_i} to find the best candidate grid point. Conceptually, the coarse search aims to find
\begin{align}
    \hat{i} = \argmax_{i\in[G]}\; |q_i|^2,
\end{align}
where we notice again that the search is always over the full grid.
Although the search is based on a coarse grid, the joint grid size $G$ is still too large to globally search all 5-D parameters for the considered wideband XL-MIMO channels.  We exploit the nested structure of $q_i$ to obtain an initial estimate for each parameter group separately, followed by an alternating maximization to find the joint optimal estimate, which then serves as the starting point for the continuous-domain refinement.  

We first show the separable structure of the atom $\phiv(\xiv)$.   
Let $\mathcal{A} \in \bC^{M_{\rm rf}\times N \times T}$ be the 3-D array expansion of $\phiv(\xiv)$, i.e.,  $\phiv(\xiv) = \vec(\mathcal{A})$. We note that the entries of $\mathcal{A}$ for a single path take the factored form\footnote{In this paper, we denote 3-D arrays by calligraphic uppercase letters, with three subscripts representing the indices of each dimension. }
\begin{align} \label{eq:atom3}
  \mathcal{A}_{m,n,t} =
  g_n(\tau) \underbrace{[\Wm_t^\herm \av(\theta^{\rm a},\theta^{\rm e},r)]_m}_{\triangleq v_{m,t}(\theta^{\rm a},\theta^{\rm e},r)}
  \underbrace{\bv^\herm (\psi)\fv_t}_{\triangleq u_t(\psi)},
\end{align}
where each of the three factors depends on a \emph{disjoint} subset of the 5-D parameters.
Using \eqref{eq:atom3}, we show the nested structure of $q_i$.
Let $\mathcal{R}\in \bC^{M_{\rm rf}\times N\times T}$ be the 3-D residual array reshaped from the Bayesian residual $\rv_\bS$, i.e., $\rv_\bS =\vec(\mathcal{R})$.
The inner product $\phiv_i^\herm \rv_\bS$ can be decomposed into nested partial sums
\begin{align}\label{eq:nested}
  \phiv_i^\herm \rv_\bS \!=\!\! \sum_{t=1}^{T} \! u_t^*(\psi_i) \! \sum_{m=1}^{M_{\rm rf}} \!  v_{m,t}^*(\theta^{\rm a}_i, \theta^{\rm e}_i,r_i)  \!  \underbrace{\sum_{n=1}^{N} \! g_n^*(\tau_i)\mathcal{R}_{m,n,t}}_{\triangleq\mathcal{Z}^{\tau}_{m,\ell,t}},
\end{align}
where $\mathcal{Z}^{\tau}\in \bC^{M_{\rm rf}\times G_\tau\times T}$, and each layer of summation eliminates one index dimension
and involves only the corresponding parameter.
Thus, we can design a multi-step search procedure that identifies good candidates of grid points without evaluating all $G$ joint grid points. We process the three factors in \eqref{eq:nested} sequentially, from the innermost sum to the outermost, each step eliminating one index dimension and narrowing the search space.

\subsubsection{Delay Transform (eliminating subcarrier index $n$)}
The innermost sum in \eqref{eq:nested} involves only the delay parameter $\tau$ and subcarrier index $n$. We can evaluate this sum for all $G_\tau$ delay candidates before considering any BS or UE angular parameters. Concretely, for each time slot $t$, we compute the matrix product\footnote{Adopting the MATLAB convention, we employ a colon to represent all indices in a given dimension.}
\begin{equation}
  \mathcal{Z}^{\tau}_{:,:,t} = \mathcal{R}_{:,:,t}\Em
  \in\bC^{M_{\rm rf} \times G_\tau},
  \label{eq:Zdef}
\end{equation}
where $[\Em]_{n,\ell} = e^{j2\pi(n-1)\Delta f \tau_\ell}$ is the $N \times G_\tau$ delay transform matrix evaluated on the delay grid $\{\tau_\ell\}_{\ell=1}^{G_\tau}$. After this step, the subcarrier index $n$ has been absorbed.
The cost is $\mathcal{O}(M G_\tau )$, independent of the BS and UE grid sizes.

\subsubsection{BS Near-Field Codebook Projection and Search (eliminating RF index $m$)}\label{sec:BS_projection}  
Now, we handle the remaining middle sum $\sum_m v^*_{m,t} \mathcal{Z}^\tau_{m,\ell,t}$ in \eqref{eq:nested}. Defining the post-combiner BS steering vector $\vv_t(\theta_\kappa^{\rm a}, \theta_\kappa^{\rm e}, r_\kappa) \triangleq\Wm^\herm_t \av(\theta_\kappa^{\rm a}, \theta_\kappa^{\rm e}, r_\kappa)$ for $\kappa$-th entry of $G_{\rm bs} \triangleq G_{\theta^{\rm a}} G_{\theta^{\rm e}} G_r$ BS angle-distance grid points and $t$-th time slot, the inner product of the $\kappa$-th BS atom with the delay-transformed residual at delay $\ell$ and slot $t$ is given as
\begin{align}\label{eq:Z_bs}
    c^{\rm bs}_{\ell, t}(\kappa) = \vv^\herm_t(\theta_\kappa^{\rm a}, \theta_\kappa^{\rm e}, r_\kappa) \mathcal{Z}^\tau_{:,\ell,t},
\end{align}
where with a slight abuse of notation in $c^{\rm bs}_{\ell, t}(\kappa)$ we directly use the grid-point index $\kappa$ as the argument for brevity.

At this point, we face two difficulties. First, evaluating \eqref{eq:Z_bs} for all $G_{\rm bs}\times G_\tau$ pairs $(\kappa,\ell)$ is very costly for large arrays. Second, the outermost sum in \eqref{eq:nested} involves the UE angle $\psi$, which is still unknown. Therefore, we cannot directly compute the coherent score $|\sum_t u^*_t(\psi)c_{\ell,t}^{\rm bs}(\kappa\in [G_{\rm bs}])|$ without searching over $\psi$, which would reintroduce undesired multiplicative cost. We address both difficulties together. For the first, we observe that $G_{\rm bs}$ grid points span $G_{\rm ang} \triangleq G_{\theta^{\rm a}}G_{\theta^{\rm e}}$ angular directions, each with $G_r$ distance samples. Most angular directions contain no scatterer, so their $G_r$ distance candidates are wasted evaluations. Therefore, we perform an \emph{angular prescreening}: for each angular direction $j\in[G_{\rm ang}]$, we select a single representative atom with the largest distance $r$ that best approximates the far-field response. We evaluate only this representative atom against $\mathcal{Z}^\tau$. To tackle the second difficulty, we eliminate the dependence on the unknown $\psi$ by using a \emph{non-coherent} combination across time slots. Specifically, since $c^{\rm bs}_{\ell,t}(\kappa)$ contains the unknown UE factor $u_t(\psi)$ as a common scalar multiplier, taking the squared modulus $|c^{\rm bs}_{\ell,t}(\kappa)|^2$ removes its phase. Then, summing the squared modulus over all slots accumulates signal energy regardless of $\psi$, so that the ranking of BS angular directions is insensitive to the unknown UE angle $\psi$, which justifies searching BS parameters before the UE angle.
The non-coherent angular score is given as
\begin{align}\label{eq:prescr}
  c_j^{\rm ang} = \max_\ell \; \sum_{t=1}^{T}
  \left|\vv_t^\herm(\kappa^{\rm far}_j) \mathcal{Z}^\tau_{:,\ell,t}\right|^2,   \quad j \in [G_{\rm ang}],
\end{align}
where $\kappa^{\rm far}_j$ is the representative atom index for direction $j$. 

We retain the top-$J$ directions with the largest $c^{\rm ang}_j$. For these $J$ directions only, we further evaluate all $G_r$ distance candidates using all near-field grid points, and pick the best (BS angle-distance, delay) pair $(\hat{\kappa}, \hat{\ell})$ as 
\begin{align}\label{eq:bstau}
    (\hat{\kappa}, \hat{\ell}) = \argmax_{\kappa \in \bI_J,\; \ell \in [G_\tau]}   \;  \sum_{t=1}^{T}
    \left| \vv_t^\herm (\kappa)
    \mathcal{Z}^\tau_{:,\ell,t} \right|^2,
\end{align}
where $\bI_J$ is the set of all grid point indices belonging to the top-$J$ directions with $|\bI_J|=JG_r$. 
The total cost of this step is $\mathcal{O}((G_{\rm ang} + JG_r) M_{\rm rf} G_\tau T)$, where the first term corresponds to the angular prescreening in \eqref{eq:prescr} and the second to the distance and delay search within the top-$J$ directions in \eqref{eq:bstau}.

\subsubsection{UE Angle Search (eliminating slot index $t$)}
Given $(\hat{\kappa}, \hat{\ell})$, we find the remaining unknown UE angle $\psi$. The full coherent score can now be evaluated because the BS and delay dimensions have been resolved.  
The optimal UE angle index is obtained by maximizing the coherent score
\begin{align}\label{eq:psisearch}
  \hat{\zeta} = \argmax_{\zeta \in [G_\psi]}\;
  \left|\sum_{t=1}^{T} u_t^*(\psi_\zeta) \vv_t^\herm (\hat{\kappa}) \mathcal{Z}^\tau_{:,\hat{\ell},t}\right|^2,
\end{align}
which amounts to a single matrix-vector product at negligible cost $\mathcal{O}(G_\psi T)$.

\subsubsection{Alternating On-Grid Refinement}
The sequential initialization may not be jointly optimal because the angular prescreening ignores the distance and the non-coherent combination discards the phase information of the UE factor. We improve it by running several rounds of cyclic coordinate maximization over the coarse grids, where each round updates one parameter group while holding the others fixed:
\begin{subequations}\label{eq:alt}
\begin{align}
  \hat{\ell} &\leftarrow
    \argmax_{\ell\in [G_\tau]}\;
    \bigl|\textstyle\sum_t u_t^*(\hat{\zeta}) \vv_t^\herm(\hat{\kappa}) \mathcal{Z}^\tau_{:,\ell,t}\bigr|^2, \label{eq:alt_tau}\\
    \hat{\zeta} &\leftarrow
    \argmax_{\zeta\in [G_\psi]}\; \bigl|\textstyle\sum_t u_t^*(\zeta) \vv_t^\herm(\hat{\kappa})
    \mathcal{Z}^\tau_{:,\hat{\ell},t}\bigr|^2, \label{eq:alt_psi}\\
    \hat{\kappa} &\leftarrow
    \argmax_{\kappa\in [G_{\rm bs}]}\; \bigl|\textstyle\sum_t u_t^*(\hat{\zeta})
    \vv_t^\herm(\kappa) \mathcal{Z}^\tau_{:,\hat{\ell},t}\bigr|^2. \label{eq:alt_bs}
\end{align}
\end{subequations}
Finally, the coarse grid search yields an estimate with grid point index $\hat{i}\in [G]$ and its position $\xiv^{\rm g}_{\hat{i}} = [\tau_{\hat{\ell}},\psi_{\hat{\zeta}},\theta^{\rm a}_{\hat{\kappa}},\theta^{\rm e}_{\hat{\kappa}},r_{\hat{\kappa}}]^\transp$.
Each update is a one-dimensional search and costs at most $\mathcal{O}(\max(G_\tau, G_\psi, G_{\rm bs}) M_{\rm rf} T)$. Note that the alternating updates in \eqref{eq:alt} use the coherent score since a UE angle estimate $\hat{\zeta}$ is now available.


\subsection{Off-Grid Damped Newton Refinement}\label{sec:newton}
The estimated $\xiv^{\rm g}_{\hat{i}}$ from coarse grid search lies on the discrete grid and thus suffers from grid quantization error. We refine $\xiv^{\rm g}_{\hat{i}}$ in the continuous domain by maximizing the evidence ratio $Q(\xiv) \triangleq |q(\xiv)|^2/s(\xiv)$ using Newton steps along each coordinate.
We use coordinate-wise Newton refinement rather than a full joint Newton update to avoid computing the full Hessian with mixed derivatives and improve numerical stability for the non-concave evidence ratio. Moreover, based on the fact that atom $\phiv(\xiv)$ can be factorized into three independent terms in \eqref{eq:atom3},    each coordinate update only requires recomputing the corresponding atom factor, while the other factors are cached and reused to reduce complexity.

For a scalar coordinate $\xi\in \{\tau, \psi, \theta^{\rm a},\theta^{\rm e}, r\}$ at the current estimate $\hat{\xi}$, we calculate the first and second derivatives $Q_\xi(\hat{\xi})$ and $Q_{\xi\xi}(\hat{\xi})$.\footnote{
In this paper, the first- and second-order derivatives of $f(x)$ are denoted by $f'(x)/f''(x)$ or $f_x/f_{xx}$ interchangeably. For variables already bearing subscripts, the prime notation is preferred to avoid subscript clutter.
}   
The Newton increment $\Delta \xi = -Q_{\xi}/Q_{\xi\xi}$ maximizes the local quadratic model, which is used for updating $\hat{\xi}$ only when it is an ascent direction, i.e., the local curvature is concave with $Q_{\xi\xi} < 0$. Rather than directly accepting the full Newton step, we adopt a backtracking line search to control the step size with a backtracking parameter $\nu\in(0,1]$ to improve robustness in the non-convex objective. Starting from $\nu = 1$, the updated point 
\begin{align}\label{eq:newton_step}
    \xi^+ = \hat{\xi} +  \nu \Delta \xi,
\end{align}
is accepted only if it strictly increases the evidence ratio, i.e., $Q(\xi^+)>Q(\hat{\xi})$; otherwise the step is shrunk with $\nu \gets \delta  \nu $ with $\delta\in(0,1)$, and retried up to a maximum number of backtracking trials $N_{\rm bt}$. We cycle through all five coordinates and repeat for $N_{\rm ref}$ passes, where the distance coordinate $r$ is refined in the $\rho = 1/r$ domain, since the near-field phase in $\rho$ is approximately affine.  
Because every accepted step strictly increases $Q$, which equivalently strictly decreases the negative marginal likelihood, the refinement never lowers the evidence and therefore preserves the monotonic convergence guarantee of the sequential SBL. 
The calculation of $Q_{\xi}$ and $Q_{\xi\xi}$ in closed-form is provided in Appendix~\ref{sec:newton_detail}.


\subsection{Active Atom Update}\label{sec:active_update}
Since the optimal position $\xiv_i^{\rm act}$ and hyperparameter $\gamma_i^{\rm act}$ of each active atom $i\in[\bar{S}]$ depend on the remaining atoms in the active set $\bar{\bS}$, we need to update the active atoms after each new atom addition. 
In order to update the $i$-th active atom in $\bar{\bS}$, we need to re-evaluate its contribution under the absence of atom $i$ and the presence of all other active atoms. Specifically, we temporarily remove atom $i$ from the model and form the leave-one-out (LOO) active set $\bar{\bS}_{-i}\triangleq \bar{\bS} \setminus\{(\xiv^{\rm act}_i, \gamma_i^{\rm act})\}$ and the corresponding LOO evidence ratio $Q_{-i}(\xiv)=|q_{-i}(\xiv)|^2/s_{-i}(\xiv)$, where the LOO quality and sparsity factors are given as 
\begin{align}
    q_{-i}(\xiv) = \phiv^\herm(\xiv)\Cm_{-i}^{-1}\yv, \quad
    s_{-i}(\xiv) = \phiv^\herm(\xiv)\Cm_{-i}^{-1}\phiv(\xiv),
\end{align}
with $\Cm_{-i} \triangleq \Cm_{\bar{\bS}_{-i}}$. 
We first refine its position by maximizing the LOO evidence ratio $Q_{-i}(\xiv)$ around $\xiv_i^{\rm act}$ using Newton steps in Section~\ref{sec:newton} and obtain new position $\xiv^{\rm new}_i$, allowing the active atoms to continuously ``drift'' toward their true parameters as the active set evolves. Having  $\xiv_i^{\rm new}$, we then calculate new LOO quality factor $q_{-i}^{\rm new}\triangleq q_{-i}(\xiv_i^{\rm new})$ and sparsity factor $s_{-i}^{\rm new} \triangleq s_{-i}(\xiv_i^{\rm new})$. If $|q_{-i}^{\rm new}|^2 > s_{-i}^{\rm new}$, we calculate new hyperparameter $\gamma_i^{\rm new}$ using \eqref{eq:gamma_opt} and update the $i$-th atom in active set as $\xiv_i^{\rm act}\gets \xiv_i^{\rm new}, \gamma_i^{\rm act}\gets \gamma_i^{\rm new}$; otherwise, atom $i$ is regarded as irrelevant and is pruned from the active set, i.e., $\bar{\bS} \gets \bar{\bS}_{-i}$. This automatic pruning mechanism determines the model order without requiring the number of paths $L$ to be specified.

Each atom update requires re-calculating $M\times M$ matrix $\Cm_{-i}^{-1}$. Following \eqref{eq:woodbury_C}, we can calculate $q_{-i}$ and $s_{-i}$ using \eqref{eq:q_mu} and \eqref{eq:s_i}, and the dimension of matrix inversion is reduced to the active-model size, but the total complexity due to this repeated matrix inversion could still be high under large $\bar{S}$. To further reduce complexity, the LOO posterior $\muv_{-i} \triangleq \muv_{\bar{\bS}_{-i}}, \Sigmam_{-i} \triangleq \Sigmam_{\bar{\bS}_{-i}}$ can be efficiently calculated from the current full posterior $\muv_{\bar{\bS}}, \Sigmam_{\bar{\bS}}$ via the Schur complement without re-inversion: 
\begin{subequations} \label{eq:loo}
\begin{align}
  \Sigmam_{-i} &= [\Sigmam_{\bar{\bS}}]_{\bar{i},\bar{i}}
    - [\Sigmam_{\bar{\bS}}]_{\bar{i},i} [\Sigmam_{\bar{\bS}}]_{i,\bar{i}} /
    [\Sigmam_{\bar{\bS}}]_{i,i},\label{eq:loo_sigma}\\ 
  \muv_{-i} &= [\muv_{\bar{\bS}}]_{\bar{i}}
    - [\Sigmam_{\bar{\bS}}]_{\bar{i},i} [\muv_{\bar{\bS}}]_i / 
    [\Sigmam_{\bar{\bS}}]_{i,i},\label{eq:loo_mu}
\end{align}
\end{subequations}
where $\bar{i} = [\bar{S}] \setminus \{i\}$ indexes all elements in $[\bar{S}]$ except $i$. 

After each atom update, we also need to update the full posterior. If the $i$-th atom is deleted, the new full posterior is simply the LOO posterior, i.e., $\Sigmam^{\rm new}_{\bar{\bS}} = \Sigmam_{-i}, \muv^{\rm new}_{\bar{\bS}} = \muv_{-i}$. Otherwise, the new full posterior can also be efficiently calculated using block-inverse update without matrix inversion:
\begin{subequations}\label{eq:full_posterior}
\begin{align}
    \Sigmam^{\rm new}_{\bar{\bS}} &= \begin{bmatrix}
        \Sigmam_{-i} + \chi^{-1}\Sigmam_{-i} \cv \cv^\herm \Sigmam_{-i} & -\chi^{-1}\Sigmam_{-i} \cv \\
        -\chi^{-1}\cv^\herm \Sigmam_{-i} & \chi^{-1}
    \end{bmatrix}, \label{eq:new_sigma}\\
    \muv^{\rm new}_{\bar{\bS}} &= \begin{bmatrix}
        \muv_{-i} - \mu_i^{\rm new}\Sigmam_{-i}\cv \\
        \mu_i^{\rm new}
    \end{bmatrix}, \label{eq:new_mu}
\end{align}
\end{subequations}
where $\quad \cv = N_0^{-1}\Phim^\herm_{-i}\phiv^{\rm new}_i \in \bC^{\bar{S}-1}$, $\chi = N_0^{-1}\|\phiv_i^{\rm new}\|^2 + 1/\gamma_i^{\rm new} - \cv^\herm \Sigmam_{-i}\cv$ with $\phiv_i^{\rm new} \triangleq \phiv(\xiv_i^{\rm new})$, $\Phim_{-i}\triangleq \Phim_{\bar{\bS}_{-i}}$, and 
\begin{align} 
    \mu_i^{\rm new} &= \chi^{-1}\left(N_0^{-1}(\phiv^{\rm new}_i)^\herm \yv - \cv^\herm \muv_{-i}\right)\\ &=\frac{q_{-i}^{\rm new}}{(\gamma_i^{\rm new})^{-1} + s_{-i}^{\rm new}}. \label{eq:mu_new_qs}
\end{align}    
The updated atom is always appended to the LOO model under the posterior update in \eqref{eq:full_posterior}. To restore the original active atom ordering for next LOO calculation, the resulting posterior quantities $\Sigmam^{\rm new}_{\bar{\bS}}$ and $\muv^{\rm new}_{\bar{\bS}}$ are then permuted back so that the updated atom occupies its original active index.
It is noticed that after adding a new atom in the active set, the full posterior can also be calculated using efficient block update in \eqref{eq:full_posterior}. 
The derivation details of matrix-inversion-free posterior updates in \eqref{eq:loo} and \eqref{eq:full_posterior} are given in Appendix~\ref{sec:rank-one}.   

\subsection{Complete Algorithm}
The proposed sequential SBL-based algorithm contains a main iteration to manage the active set $\bar{\bS}$. In each main iteration, adding a new atom and updating active atoms are applied in turn. To add a new atom, the decoupled coarse search (Section~\ref{sec:on-grid}) is executed on the current Bayesian residual $\rv_{\bar{\bS}}$, which yields a chosen grid point $\hat{i}$ with position $\xiv^{\rm g}_{\hat{i}}$. Followed by Newton refinement (Section~\ref{sec:newton}) yields a refined position $\xiv^{\rm off}_{\hat{i}}$ and optimal hyperparameter $\gamma^\star_{\hat{i}}$ given by \eqref{eq:gamma_opt}. 
If the resulting evidence increment satisfies $-\Delta\mathcal{L}^\star > \epsilon$ with $\epsilon>0$ being a small threshold, the candidate is added to the active set, i.e., $\bar{\bS} \gets \bar{\bS}\cup\{(\xiv^{\rm off}_{\hat{i}}, \gamma^\star_{\hat{i}})\}$. 
After adding a new atom in each main iteration, the positions and hyperparameters of all active atoms are updated, and irrelevant atoms would be pruned (Section~\ref{sec:active_update}). 
The algorithm terminates when no action improves the evidence, and returns with active atom positions $\xiv_{\bar{\bS}} \triangleq \{\xiv_i^{\rm act} \in \bar{\bS} \mid i\in[\bar{S}]\}$ and corresponding posterior means $\muv_{\bar{\bS}}$. The full channel is reconstructed using the multipath channel model in \eqref{eq:channel} by treating the posterior means as the estimated channel complex gain $\widehat{\alphav}=\muv_{\bar{\bS}}$. 
The complete algorithm is given in Algorithm~\ref{alg:main}.

\begin{algorithm}[t]
\caption{Decoupled Off-Grid Sequential SBL}
\label{alg:main}
\begin{algorithmic}[1]
\REQUIRE Observations $\yv$, pilots
  $\{\fv_t, \Wm_t\}_{t=1}^T$,  noise power $N_0$
\STATE Initialize $\bar{\bS} \leftarrow \emptyset$, $\gammav \leftarrow \mathbf{0}$, $J$ = 20, $\epsilon = 0.001$

\REPEAT
  \STATE \textbf{--- \textit{Add new atom} ---}
  \STATE Compute residual $\rv_{\bar{\bS}}$ and reshape it to $\mathcal{R}$
  \STATE \textbf{Delay transform:} $\mathcal{Z}^\tau_{:,:,t}
    = \mathcal{R}_{:,:,t}\mathbf{E}$, $\forall t$
    \hfill (\ref{eq:Zdef})
  \STATE \textbf{Angular prescreening:} find top-$J$ directions
    \hfill (\ref{eq:prescr})
  \STATE \textbf{Distance search:} within top-$J$ directions \hfill \eqref{eq:bstau}
  \STATE \textbf{UE angle search:} search $\psi$
    \hfill (\ref{eq:psisearch})
  \STATE \textbf{Alternating refinement:} several (e.g., 3) rounds
    \hfill (\ref{eq:alt})
  \STATE \textbf{Newton off-grid refinement:} refine $\xiv^{\rm g}_{\hat{i}}\to \xiv^{\rm off}_{\hat{i}}$ 
    \hfill (\ref{eq:newton_step})
  \STATE Evaluate evidence increment $q_{\hat{i}}, s_{\hat{i}} \to Q_{\hat{i}} \to
    \Delta\mathcal{L}^\star$ \hfill \eqref{eq:function_Q}
  \IF{$-\Delta\mathcal{L}^\star > \epsilon$}
    \STATE Calculate hyperparameter: $\gamma_{\hat{i}}^\star \gets (|q_{\hat{i}}|^2 - s_{\hat{i}})/s_{\hat{i}}^2$
    \STATE Add new atom: $\bar{\bS} \gets \bar{\bS} \cup \{(\xiv^{\rm off}_{\hat{i}},\gamma^\star_{\hat{i}})\}$
    \STATE Rank-one update full posterior \hfill \eqref{eq:full_posterior}
  \ENDIF
  \STATE \textbf{--- \textit{Update and delete active atoms} ---}
  \FOR{each $i \in [\bar{S}]$}
    \STATE Compute LOO posterior \hfill \eqref{eq:loo}
    \STATE Newton-refine $Q_{-i}(\xiv)$ around $\xiv^{\rm act}_i \to \xiv^{\rm new}_i$ \hfill \eqref{eq:newton_step}
    \IF{$|q_{-i}^{\rm new}|^2 > s_{-i}^{\rm new}$} 
    \STATE{$\gamma_i^{\rm new} \gets (|q_{-i}^{\rm new}|^2 - s_{-i}^{\rm new})/(s_{-i}^{\rm new})^2$}
    \STATE Update atom: $\xiv^{\rm act}_i \gets \xiv^{\rm new}_i, \gamma^{\rm act}_i \gets \gamma^{\rm new}_i$
    \STATE Rank-one update full posterior \hfill \eqref{eq:full_posterior}
    \ELSE 
    \STATE Delete atom: $\bar{\bS} \gets \bar{\bS}_{-i}$
    \STATE Update full posterior: $\Sigmam_{\bar{\bS}} \gets \Sigmam_{-i}, \muv_{\bar{\bS}}\gets \muv_{-i}$ 
    \ENDIF

  \ENDFOR
\UNTIL{no action improves $\mathcal{L}$}
\STATE \textbf{return} $\{\xiv_{\bar{\bS}}, \muv_{\bar{\bS}}\}$,
  reconstruct $\{\widehat{\Hm}[n]\}$ via \eqref{eq:channel}
\end{algorithmic}
\end{algorithm}

\begin{remark}
The efficient grid-wide recursive update mechanism developed in \cite{pote2025theory} cannot be directly adopted in our case, since it requires maintaining quality and sparsity factors over the full joint grid, which is computationally and memory prohibitive due to the extremely large grid size $G$ in the considered 5-D parameter space. Moreover, such recursive updates rely on a fixed dictionary throughout the sequential search. This is why the extended gridless algorithm in \cite{pote2025theory} performs the off-grid refinement only after completing the on-grid selection stage, resulting in the same grid mismatch error at every step of the on-grid phase.  In contrast, our method refines each
accepted atom immediately and maintains only the posterior quantities of the small active model. \hfill $\lozenge$
\end{remark}

\subsection{Complexity Analysis}
We first emphasize that with the maintained posterior covariance and mean, both the LOO posterior
extraction and the insertion of a new or updated atom can be carried out by
Schur-complement and block-inverse updates. Therefore, the algorithm avoids
all explicit non-scalar  matrix inversions during the sequential updates.
Moreover, since the posterior updates and Newton refinement are performed only over the small active model with negligible complexity compared to the on-grid search, we focus on the dominant search cost.
The total cost of the decoupled coarse search is the cost summation of delay transform, BS codebook search, and UE angle search. Normally, the BS codebook size is much larger than the delay grid size and UE angle grid size, and thus the total cost is dominated by the BS prescreening operation in Section~\ref{sec:BS_projection}.
Table~\ref{tab:cost} summarizes the per-iteration dominant cost, where the typical scales are calculated using the simulation parameters given in Section~\ref{sec:simulation}.
The proposed method achieves two levels of complexity reduction
compared to classical approaches:
(i) sequential SBL avoids the $\mathcal{O}(M^2 G)$ global posterior computation required by SBL-EM; (ii) the decoupled search replaces the multiplicative joint-grid evaluation with  additive lightweight operations.

In addition to computational saving, the proposed method also reduces the memory requirement. Conventional full-grid SBL implementations require access to the full joint dictionary of size $G$, either by storing $\Phim\in \bC^{M\times G}$ explicitly or by maintaining the quality and sparsity factors for all $G$ grid points. In contrast, our decoupled approach avoids constructing the full dictionary and only stores the factorized low dimensional sub-dictionaries, reducing the memory significantly.

\begin{table}[t]\centering
\caption{Per-iteration computational cost comparison.}
\label{tab:cost}
\vspace{-2mm}
\begin{tabular}{@{}lcl@{}}
\toprule
Method & Dominant cost  & Typical scale\\
\midrule
SBL-EM (joint grid) &  $\mathcal{O}(M^2 G)$ &  $\sim10^{17}$ \\[2pt]
Seq.\ SBL (joint grid) & $\mathcal{O}(MG)$ & $\sim10^{12}$ \\[2pt] 
\textbf{Proposed (decoupled)} & $\mathcal{O}(G_{\rm ang} M_{\text{rf}} G_\tau T)$ & $\sim10^8$\\
\bottomrule
\end{tabular}
\end{table}

\subsection{Cram\'er--Rao Lower Bound}
We derive the CRLB for the MSE of the studied channel estimation problem.
Recall $\xiv_l = [\tau_l, \psi_l, \theta^{\rm a}_l, \theta^{\rm e}_l, r_l]^\transp$ and let $\thetav_l \triangleq [\Re\{\alpha_l\}, \Im\{\alpha_l\}, \xiv_l^\transp]^\transp$, the pilot observation in \eqref{eq:linear_form} is a parametric function $\yv = \muv(\thetav) + \widetilde{\nv}$ of the real-valued parameter vector $\thetav = [\thetav_1^\transp,\dots, \thetav_L^\transp]^\transp\in\bR^{7L}$, where $\muv(\thetav) \triangleq \sum^L_{l=1}\alpha_l \phiv(\xiv_l)$.
Since $\widetilde{\nv}\sim\mathcal{CN}(\mathbf{0},N_0\Id)$ and the noise power $N_0$ is known, by applying the Slepian-Bangs formula, the elements of the Fisher information matrix (FIM) can be derived as
\begin{align}
    [\Fm]_{i,j} = \frac{2}{N_0}\Re\left\{\frac{\partial\muv^\herm}{\partial\theta_i} \frac{\partial\muv}{\partial\theta_j}\right\},
\end{align}
where $\theta_i = [\thetav]_i$, $\partial \muv/ \partial \Re\{\alpha_l\} = \phiv(\xiv_l)$, $\partial \muv/ \partial \Im\{\alpha_l\} = j\phiv(\xiv_l)$, $ \partial \muv / \partial \xi_l = \alpha_l \phiv_{\xi_l}$ and  $\phiv_{\xi_l}$ is given in Appendix~\ref{sec:newton_detail}.
To obtain the CRLB for the full wideband channel vector $\hv$, we apply the Jacobian transformation $\Jm = \partial \hv / \partial \thetav^\transp \in \bC^{\widetilde{M}\times 7L}$, and the resulting normalized mean square error (NMSE) lower bound is given as 
\begin{align}
    \text{NMSE}_\text{CRLB} = \trace(\Jm \Fm^{-1} \Jm^\herm) / \|\hv\|^2_2,
\end{align}
where $\partial \hv /\partial \thetav^\transp$ can be computed in the same manner as $\partial \muv / \partial \thetav^\transp$, except that the full channel atoms are used instead of the pilot-domain atoms.


\section{Wideband Cooperative Hybrid Precoding}
In this section, we consider the DL hybrid precoding design for multi-user data transmission enabled by estimated channel parameters. It was theoretically proved in \cite{hellings2011inseparability} that parallel MIMO broadcast channels with linear transceivers are not always separable. This means that designing the precoder independently for each subcarrier could be suboptimal. However, \cite{hellings2011inseparability} did not provide a constructive precoder design, which is challenging under the hardware constraints of wideband hybrid XL-MIMO systems. Motivated by this, we consider the novel subcarrier-cooperative transmission scheme, which allows each data symbol to be spread over several subchannels corresponding to the case of joint encoding over the parallel channels.

After channel acquisition, we consider $K$ users to serve in DL transmission. By adding a subscript $k$ in (\ref{eq:channel}), we  denote the channel of user $k$ as
\begin{align}\label{eq:channel_k}
    \Hm_k[n] = \sum\nolimits^L_{l=1} \alpha_{k,l}  g_n(\tau_{k,l}) \av(\theta^{\rm a}_{k,l}, \theta^{\rm e}_{k,l}, r_{k,l}) \bv^\herm(\psi_{k,l}),
\end{align}
where all multipath parameters have been estimated by the proposed SBL algorithm.
Based on the channel reciprocity, the DL MIMO channel from the BS to the $k$-th user at subcarrier $n$ is denoted by $ \Hm_k^\herm[n] \in \bC^{M_{\rm ue}\times M_{\rm bs}}$. Then, the overall wideband channel is written as 
\begin{align}
\Hm_k^\herm   \triangleq    {\sf blkdiag} (\Hm_k^\herm[1] ,\dots, \Hm_k^\herm[N] )  \in \bC^{M_{\rm ue}N\times M_{\rm bs}N}.
\end{align}
The subcarrier-cooperative signal transmitted by the BS can be expressed as
\begin{align}
\mathbf{x}=\wideparen{\mathbf{W}}_{\rm RF}  \sum\nolimits_{k=1}^K \mathbf{W}_{{\rm BB},k}\mathbf{s}_k  = \wideparen{\mathbf{W}}_{\rm RF}\mathbf{W}_{{\rm BB}} \mathbf{s}\in \bC^{M_{\rm bs} N}, 
\end{align}
where $ \wideparen{\mathbf{W}}_{\rm RF} \triangleq \mathbf{I}_N \otimes \mathbf{W}_{\rm RF}   $ is the overall analog precoder, with $\mathbf{W}_{\rm RF} \in \bC^{M_{\rm bs}\times M_{\rm rf}}$ being the frequency-independent analog precoder common for all subcarriers. For the considered sub-connected structure, we have $\mathbf{W}_{\rm RF} \triangleq {\sf blkdiag} ( \mathbf{w}_{{\rm RF},1} ,\dots, \mathbf{w}_{{\rm RF},M_{\rm rf}} )  $ with unit-modulus constraint $| [\mathbf{w}_{{\rm RF},m_{\rm rf}}]_i |=1/\sqrt{\beta}$.  $\mathbf W_{\rm BB}\triangleq [\mathbf W_{{\rm BB},1},\ldots,\mathbf W_{{\rm BB},K}]$ is the overall BS digital precoder and $\mathbf{s} \triangleq [\mathbf{s}_1^\transp,\dots,\mathbf{s}_K^\transp]^\transp$ is the transmitted symbol with $\mathbb{E}[\mathbf{s} \mathbf{s}^\herm]=\mathbf{I}$, where $\mathbf{W}_{{\rm BB},k} \in \bC^{M_{\rm rf}N\times S_{k}}$ and $\mathbf{s}_k  \in \bC^{S_{k}}$ are  subcarrier-cooperative digital precoder and transmitted data symbol for user $k$, respectively\footnote{It is worth noting that by letting $\mathbf{W}_{{\rm BB},k}$ be of a block-diagonal structure, i.e., $\mathbf{W}_{{\rm BB},k}= {\sf blkdiag} ( \mathbf{w}_{{\rm BB},k}[1],\dots, \mathbf{w}_{{\rm BB},k}[N]) $, the considered transmission scheme will naturally degrade to the conventional subcarrier-independent transmission scheme, e.g., in \cite{yuan2023alternating}.}. Moreover, the number of streams should satisfy $S_k\leq N$ due to a single RF chain at each user. We assume $S_k=N$ for simplicity.  

The received signal at user $k$ with a frequency-independent analog combiner $\mathbf{f}_k   \in \bC^{ M_{\rm ue} } $, $| [\mathbf{f}_k]_i|=1/\sqrt{M_{\rm ue}}$,  is given by
\begin{align}
{\mathbf{y}}_k = \wideparen{\mathbf{F}}_{{\rm RF},k}  \Hm_k^\herm \mathbf{x}   +  \wideparen{\mathbf{F}}_{{\rm RF},k}  \mathbf{n}_k,
\end{align}
where $\wideparen{\mathbf{F}}_{{\rm RF},k} \triangleq \mathbf{I}_N \otimes \mathbf{f}_k^\herm$ and $\mathbf{n}_k   \sim\mathcal{CN}(\mathbf{0},\sigma_{\rm d}^2 \Id_{M_{\rm ue} N}) $ are   $N$-subcarrier overall combiner and noise, respectively, with DL noise power $\sigma^2_{\rm d}$. Importantly, we have $\wideparen{\mathbf{F}}_{{\rm RF},k} \wideparen{\mathbf{F}}_{{\rm RF},k}^\herm = \Id_{M_{\rm ue}N}$.

For notation brevity, define $\mathbf{Z}_{k, u}   \triangleq \wideparen{\mathbf{F}}_{\mathrm{RF}, k} \mathbf{H}_k^{ {\herm }} \wideparen{\mathbf{W}}_{\mathrm{RF}} \mathbf{W}_{\mathrm{BB}, u}$.  Assume that user $k$ jointly decodes $\mathbf{s}_k $ from the stacked observation ${\mathbf{y}}_k$.  By treating interference as noise, the achievable rate is given by
\begin{align}\label{rate_experssion}
R_k =   \log \!  \left|\mathbf{I}_N   \!+\!     \mathbf{Z}_{k, k}^{\herm}
\big(\sum\nolimits_{u \neq k}  \!  \mathbf{Z}_{k, u} \mathbf{Z}_{k, u}^{\herm}  \! +\!    \sigma_{\mathrm{d}}^2 \mathbf{I}_N   \big)^{-1} 
\mathbf{Z}_{k, k}\right|,
\end{align}
and the sum user rate maximization problem is formulated as
\begin{subequations}\label{optimization_problem}
	\begin{align} \label{obj}
		&  \mathop {\max }\limits_{       \mathbf{W}_{\rm RF} ,
       \{ {\mathbf{W}_{{\rm BB},k}} \},          \{ \mathbf{f}_{k    } \}     } \;  \;  \;  \sum\nolimits_{k=1}^K        R_{k}       \\
\label{power_constraint}
&\text { s.t. }  \quad  \sum\nolimits_{k = 1}^{{K}} {   \left\|  \wideparen{\mathbf{W}}_{{\rm RF}}\mathbf{W}_{{\rm BB},k}  \right\|_{\sf F}^2  \le   P_{\rm d} },    \\\label{bs_phase_constrinat}
&       \quad\qquad    \left|   \left[ \mathbf{w}_{{{\rm RF},m_{\rm rf}}}  \right]_i \right| = 1/\sqrt{\beta}    , \;\;  i \in [\beta],  m_{\rm rf} \in [M_{\rm rf}], \\ \label{ue_phase_constrinat}
&       \quad\qquad    \left|   \left[ \mathbf{f}_{k}  \right]_i \right| = 1/\sqrt{M_{\rm ue}}  , \;\; i\in [M_{\rm ue}],  k\in[K],
	\end{align}
\end{subequations}
where $P_{\rm d}$ is the DL total transmit power. For the power constraint (\ref{power_constraint}), since $\wideparen{\mathbf{W}}_{{\rm RF}}^\herm  \wideparen{\mathbf{W}}_{{\rm RF}}=\mathbf{I}$, we equivalently have
\begin{align}\label{Wbb_power}
 \sum_{k = 1}^{{K}}    \left\|   \wideparen{\mathbf{W}}_{{\rm RF}}\mathbf{W}_{{\rm BB},k}  \right\|_{\sf F}^2 
 \!  = \! \sum_{k = 1}^{{K}}    \left\|   \mathbf{W}_{{\rm BB},k}  \right\|_{\sf F}^2
 \! =\!   \left\|   \mathbf{W}_{{\rm BB}}  \right\|_{\sf F}^2  
   \le  \!   P_{\rm d} .
\end{align}

\subsection{The Alternating Optimization Algorithm} \label{alternating_section}
We first propose an alternating optimization algorithm supported by the estimated channel matrix $\Hm_k$.
To tackle the non-convex problem (\ref{optimization_problem}), we first apply the matrix quadratic transform and matrix Lagrangian dual transform \cite{shen2019optimization},  so that the fractional objective function (\ref{obj}) can be recast to
\begin{align}\label{recast}
\begin{aligned}
	&f_o  \left(     {\mathbf{W}}_{{\rm RF}} ,
	\mathbf{W}_{{\rm BB},k} , \mathbf{f}_{k} , \mathbf{Y}_k, \boldsymbol{\Gamma}_k        \right)   
	 =     \sum\nolimits_{k=1}^{K}           \Big \{        \log \left|\mathbf{I}+\boldsymbol{\Gamma}_k\right| \\
	&- {\tr}\left(\boldsymbol{\Gamma}_k\right) 
	 +2 \Re\left\{  {\tr}    \left(\left(\mathbf{I}+\boldsymbol{\Gamma}_k\right) \mathbf{Z}_{k, k}^{  { \herm }} \mathbf{Y}_k\right)\right\} \\
	& - {\tr}\left(\left(\mathbf{I}+\boldsymbol{\Gamma}_k\right) \mathbf{Y}_k^{  { \herm }}\left(
	\sum\nolimits_{u=1}^K  \!    \mathbf{Z}_{k, u} \mathbf{Z}_{k, u}^{ {\herm }}+\sigma_{\mathrm{d}}^2 \mathbf{I}_N\right) \mathbf{Y}_k\right)  \Big\}, 
\end{aligned}
\end{align}
where  $ \boldsymbol{\Gamma}_k \in \bC^{N\times N}  $ (Hermitian positive  semi-definite matrix) and  $\mathbf{Y}_k\in \bC^{N\times N}$  are two auxiliary variables.
Given $   \wideparen{\mathbf{W}}_{{\rm RF}}$ ,
${\mathbf{W}_{{\rm BB},k}} $,  and $ \wideparen{ \mathbf{F}}_{{\rm RF},k}  $, by letting first-order derivative equal to zero, the optimal solutions of the two auxiliary variables   $ \boldsymbol{\Gamma}_k $ and  $ \mathbf{Y}_k$ are given by
\begin{align}
& \mathbf{\Gamma}_k^\star=\mathbf{Z}_{k, k}^{ {\herm }} 
\left(\sum\nolimits_{u \neq k} \mathbf{Z}_{k, u} \mathbf{Z}_{k, u}^{\mathrm{H}}+\sigma_{\mathrm{d}}^2 \mathbf{I}_N\right)^{-1} 
 \mathbf{Z}_{k, k}, \\
& \mathbf{Y}_k^\star   =\left(\sum\nolimits_{u=1}^K \mathbf{Z}_{k, u} \mathbf{Z}_{k, u}^{ {\herm }}+\sigma_{\mathrm{d}}^2 \mathbf{I}_N\right)^{-1} \mathbf{Z}_{k, k}.
\end{align}
In the following, we will solve  $   \wideparen{\mathbf{W}}_{{\rm RF}}$ ,
${\mathbf{W}_{{\rm BB},k}} $,  and $ \wideparen{ \mathbf{F}}_{{\rm RF},k}  $ in an alternating way.

\subsubsection{BS Digital Precoding}
Based on (\ref{recast}), fixing all other variables, the design of digital precoder $\mathbf{W}_{\rm BB}$ under power constraint can be  formulated as
\begin{subequations}\label{optimization_problem_digital}
	\begin{align} \label{obj_digital}
		&  \mathop {\max }\limits_{        
			\{ {\mathbf{W}_{{\rm BB},k}} \}            }  \sum_{k=1}^K
		\left[
	2\Re\{\tr(\mathbf W_{{\rm BB},k}^{\herm}\mathbf D_k)\}
-
\tr(\mathbf W_{{\rm BB},k}^{\herm}\mathbf B\mathbf W_{{\rm BB},k})
\right]      \\
		&\quad \text { s.t. }  \quad   \sum\nolimits_{k = 1}^{{K}}    \left\|   \mathbf{W}_{{\rm BB},k}  \right\|_{\sf F}^2 
	    \le      P_{\rm d}, \label{eq:power_constraint}
	\end{align}
\end{subequations}
where $\mathbf D_k
\triangleq
(    \wideparen{ \mathbf F}_{{\rm RF},k}\mathbf H_k^{\herm}   \wideparen{\mathbf W}_{\rm RF})^{\herm}
\mathbf Y_k
(\mathbf I_N+\boldsymbol\Gamma_k)$ and $\mathbf B
\triangleq
\sum\nolimits_{u=1}^K \mathbf{D}_u \mathbf{Y}_u^{\herm}      \wideparen{\mathbf{F}}_{\mathrm{RF}, u} \mathbf{H}_u^{\herm} \wideparen{\mathbf{W}}_{\mathrm{RF}}$.
 While problem (\ref{optimization_problem_digital})  can be solved by Lagrange Multipliers, the complexity of matrix inversion is prohibitive due to XL-MIMO. To avoid the heavy computational burden, fast iterative optimization under minorization-maximization (MM) principle can be used \cite{zhi2022ZF}.  Ignoring constraint (\ref{eq:power_constraint}), for the $(t+1)$-th iteration,  we have   inverse-free update as follows \cite{shen2024accelerating} 
\begin{align}
\breve{\mathbf{W}}_{\mathrm{BB}, k} =\mathbf{W}_{\mathrm{BB}, k}^{(t)}+\frac{1}{\eta}\left(\mathbf{D}_k-\mathbf{B} \mathbf{W}_{\mathrm{BB}, k}^{(t)}\right), 
\end{align}
where  $ \mathbf{W}_{\mathrm{BB}, k}^{(t)} $ is the variable value at the $t$-th iteration and $ \eta \geq \lambda_{\max }(\mathbf{B})  $.
For computational simplicity, we use $\eta=\|\mathbf{B}\|_{\sf F} $. Then, digital precoder is obtained by projecting 
$\{     \breve{\mathbf{W}}_{\mathrm{BB}, k}    \}_{k=1}^K$ onto the transmit power constraint as follows
\begin{align}
	\mathbf W_{{\rm BB},k}^{(t+1)}
	=  
	\begin{cases}
		  &\!\!\!\!\!\!   \breve{\mathbf{W}}_{\mathrm{BB}, k},   \text{         if  } 
		\sum_{u=1}^K\|    \breve{\mathbf{W}}_{\mathrm{BB}, u}     \|_{\sf F}^2\le P_{\rm d},
		\\[1ex]
		& \!\!\!\!\!\!     \sqrt{
			\dfrac{P_{\rm d}}
			{\sum_{u=1}^K\|      \breve{\mathbf{W}}_{\mathrm{BB}, u}       \|_{\sf F}^2}
		}
		\breve{\mathbf{W}}_{\mathrm{BB}, k},\; 
		\text{otherwise}.
	\end{cases}
\end{align}
Letting $I_{\rm BB}$ denote the number of MM iterations, the overall computational complexity for digital precoder is $ \mathcal{O}\left(K M_{\mathrm{rf}}^2 N^2+K N^3+I_{\mathrm{BB}} K M_{\mathrm{rf}}^2 N^3\right) $.

\subsubsection{User-Side Analog Combiner}
Similar to the derivation of (\ref{obj_digital}),  by fixing all the other variables, we can readily formulate the objective function with respect to $ \wideparen{\mathbf{F}}_{\mathrm{RF}, k} $ as a quadratic function. However, the unique challenge in the considered subcarrier-cooperative transmission is to formulate an optimization problem with $\mathbf f_k$. This means that we need extract  $\mathbf f_k$ from its  Kronecker product $\wideparen{\mathbf{F}}_{{\rm RF},k} = \mathbf{I}_N \otimes \mathbf{f}_k^\herm$.

To this end,  define $ \mathbf A_{k,u} \triangleq \mathbf H_k^\herm \wideparen{\mathbf W}_{\rm RF} \mathbf W_{{\rm BB},u}  \in \mathbb{C}^{  M_{\mathrm{ue}}N \times N}$ so that
$ 	\mathbf Z_{k,u}
=
(\mathbf I_N\otimes \mathbf f_k^\herm)\mathbf A_{k,u} $. Divide $  \mathbf A_{k,u} =  [	\mathbf A_{k,u}^\transp[1], \dots,
\mathbf A_{k,u}^\transp[N]     ]^\transp$ into $N$ parts to have $\mathbf{A}_{k, u}[n] \in \mathbb{C}^{M_{\mathrm{ue}} \times N}$.  We can then explicitly construct
\begin{align}\label{rvec_feature}
	\operatorname{rvec}(\mathbf Z_{k,u})
=
\mathbf f_k^\herm \mathbf {\breve{ \mathbf A}}_{k,u},
\end{align}
where $\operatorname{rvec}(\cdot)$ is a row-wise vectorization operator and ${\breve{ \mathbf A}}_{k,u} \triangleq  [	\mathbf A_{k,u}[1], \dots,
\mathbf A_{k,u}[N]     ] \in \mathbb{C}^{M_{\mathrm{ue}} \times N^2}$.
To convert the trace operator as shown in   (\ref{obj_digital})  into a row-vectorization operator, we need the following properties:
\begin{align}\label{trace_to_rvec}
\begin{aligned}
& \tr \left(\mathbf{Z}^{\herm} \mathbf{X}\right)=\operatorname{rvec}(\mathbf{X}) \operatorname{rvec}(\mathbf{Z})^{\herm}   , \\
&   \tr   \left(\mathbf{Z}^{\herm} \mathbf{X} \mathbf{Z}\right)=\operatorname{rvec}(\mathbf{Z})\left(\mathbf{X}^{\transp} \otimes \mathbf{I}\right) \operatorname{rvec}(\mathbf{Z})^{\herm}  .
\end{aligned}
\end{align}
Applying (\ref{rvec_feature}) and (\ref{trace_to_rvec}) to (\ref{recast}), we can  formulate the optimization problem with $\mathbf f_k$ as follows
  \begin{subequations}\label{optimization_problem_user_analog}
	\begin{align} \label{obj_user_analog}
		&  \max _{\left\{\mathbf{f}_k\right\}} \quad \sum\nolimits_{k=1}^K\left[2 \operatorname{Re}\left\{\mathbf{u}_k^{\herm} \mathbf{f}_k\right\}-\mathbf{f}_k^{\herm} \mathbf{U}_{\mathbf{f},k} \mathbf{f}_k\right]  \\
		&\; \text { s.t. }  \quad     \left|   \left[ \mathbf{f}_{k}  \right]_i \right| = 1/\sqrt{M_{\rm ue}}  , \;\;  i \in  [M_{\rm ue}],  k\in[K],
	\end{align}
\end{subequations}
where $ \mathbf{u}_k^{\herm} \triangleq \operatorname{rvec}\left(\mathbf{Y}_k\left(\mathbf{I}_N+\boldsymbol{\Gamma}_k\right)\right) \breve{\mathbf{A}}_{k, k}^{\herm}  $ and
\begin{align}
\begin{aligned}
 \mathbf{U}_{\mathbf{f},k} \!  \triangleq \!
 \sum\nolimits_{u=1}^K \! \!    \breve{\mathbf{A}}_{k, u}  \!    \left(\left[\mathbf{Y}_k\left(\mathbf{I}_N+\boldsymbol{\Gamma}_k\right) \mathbf{Y}_k^{\herm}\right]^{\transp} \!  
 \otimes \mathbf{I}_N   \!    \right) \breve{\mathbf{A}}_{k, u}^{\herm}.
\end{aligned}
\end{align}

Problem (\ref{optimization_problem_user_analog}) is a standard quadratic optimization problem with unit-modulus constraints, which can be solved with low complexity by MM algorithms \cite[(56)]{zhi2025holographic}. Let $I_{\rm f}$ be the MM iteration number. The computational complexity for designing $\mathbf{f}_k$ is $ \mathcal{O}\left(K^2 M_{\mathrm{ue}}^2 N^2+K^2 M_{\mathrm{ue}} N^3+I_{\mathrm{f}} K M_{\mathrm{ue}}^2\right) $.

\subsubsection{BS-Side Analog Precoding} Similarly, it is readily to find that (\ref{recast}) is a quadratic function of $  \wideparen{\mathbf W}_{\rm RF}$. Next, the unique technical challenge is to formulate a tractable optimization problem with respect to $   {\mathbf W}_{\rm RF} $, from its coupled Kronecker relationship in $ \mathbf Z_{k,u} $, and considering its sub-connected feature.
 
Divide $ \mathbf{W}_{\mathrm{BB}, u} = \left[\mathbf{W}_{\mathrm{BB}, u}^\transp[1], \ldots, \mathbf{W}_{\mathrm{BB}, u}^\transp[N]\right]^\transp $ into $N$ parts to have $ \mathbf{W}_{\mathrm{BB}, u}[n] \in \mathbb{C}^{M_{\mathrm{rf}} \times N} $.
Exploiting the block-diagonal feature of $ \wideparen{\mathbf{F}}_{\mathrm{RF}, k}  $, $  \mathbf{H}_k^{ {\herm }}  $, and $  \wideparen{\mathbf W}_{\rm RF}$,  the $n$-th row of $\mathbf{Z}_{k, u} $ can now be expressed as 
\begin{align}\label{Z_n_row}
\begin{aligned}
	\left[  \mathbf{Z}_{k, u}   \right]_{n,:} = \mathbf{f}_k^{\herm} \mathbf{H}_k^{\herm}[n] \mathbf{W}_{\mathrm{RF}} \mathbf{W}_{\mathrm{BB}, u}[n].
\end{aligned}
\end{align}
Based on the sub-connected structure, we define $\mathbf{w}_{\mathrm{RF}} \triangleq\left[\mathbf{w}_{\mathrm{RF}, 1}^\transp, \ldots, \mathbf{w}_{\mathrm{RF}, M_{\mathrm{rf}}}^\transp \right]^\herm   \in \mathbb{C}^{M_{\rm bs}  } $ as the optimization variables of $M_{\rm bs} $ analog phase shifters so that $ \mathbf{W}_{\mathrm{RF}}=\diag\left(\mathbf{w}_{\mathrm{RF}}^* \right)\left(\mathbf{I}_{M_{\mathrm{rf}}} \otimes \mathbf{1}_\beta\right) $, where $\mathbf{1}_\beta$ is a vector with all-ones elements. Applying the property $\mathbf{a}^\transp\diag(\mathbf{b})=\mathbf{b}^\transp \diag (\mathbf{a})$ to (\ref{Z_n_row}), we can explicitly express $\mathbf{w}_{\mathrm{RF}}$ as
\begin{align}
	\left[  \mathbf{Z}_{k, u}   \right]_{n,:} =\mathbf{w}_{\mathrm{RF}}^\herm \mathbf{C}_{k, u}[n] ,
\end{align} 
where $\mathbf{C}_{k, u}[n] \triangleq \diag    \left(\mathbf{H}_k^*[n] \mathbf{f}_k^*\right)\left(\mathbf{I}_{M_{\mathrm{rf}}} \otimes \mathbf{1}_\beta\right) \mathbf{W}_{\mathrm{BB}, u}[n]  \in \mathbb{C}^{M_{\mathrm{bs}} \times N } $.
We can further obtain
$ \operatorname{rvec}\left(\mathbf{Z}_{k, u}\right)=\mathbf{w}_{\mathrm{RF}}^\herm \breve{\mathbf{C}}_{k, u} $ with  $\breve{\mathbf{C}}_{k, u}\triangleq \left[\mathbf{C}_{k, u}[1], \ldots, \mathbf{C}_{k, u}[N]\right]  \in \mathbb{C}^{M_{\mathrm{bs}} \times N^2 }$. With (\ref{trace_to_rvec}),   the optimization problem with respect to $\mathbf{w}_{\mathrm{RF}}  $ is formulated as
\begin{subequations}\label{optimization_problem_bs_analog}
	\begin{align} \label{obj_bs_analog}
		&  \max_{\mathbf{w}_{\mathrm{RF}}     }  \quad    2 \operatorname{Re}\left\{     {\mathbf{g}}^{\herm} \mathbf{w}_{\rm RF}\right\}- \mathbf{w}_{\rm RF}^{\herm} \mathbf{G}   \mathbf{w}_{\rm RF}    \\
		&\;\text { s.t. }  \quad   \left|     \left[ \mathbf{w}_{\mathrm{RF}}  \right]_{m}    \right| = 1/\sqrt{\beta},\;  m \in [M_{\rm bs}],
	\end{align}
\end{subequations}
where $ 	\mathbf{g}^\herm  =\sum\nolimits_{k=1}^K \operatorname{rvec}\left(\mathbf{Y}_k\left(\mathbf{I}+\boldsymbol{\Gamma}_k\right)\right) \breve{\mathbf{C}}_{k, k}^{\herm} $ and 
\begin{align}
\begin{aligned}
	\mathbf{G} \!=\!  \sum\nolimits_{k=1}^K \sum\nolimits_{u=1}^K  \!\!   \breve{\mathbf{C}}_{k, u}  \!   \left(\left[\mathbf{Y}_k\left(\mathbf{I}+\boldsymbol{\Gamma}_k\right) \mathbf{Y}_k^{\herm}\right]^{\transp} \!    
\otimes \mathbf{I}\!  \right) \breve{\mathbf{C}}_{k, u}^{\herm},
\end{aligned}
\end{align}
which, again, can be  readily solved by MM algorithms. The computational complexity is $\mathcal{O}\left(K^2 M_{\mathrm{bs}}^2 N^2+I_{\mathrm{w}} M_{\mathrm{bs}}^2\right)$ with $I_{\mathrm{w}}$ being the MM iteration number. This is the heaviest part of the alternating algorithm due to the large number of $M_{\mathrm{bs}}$.


\subsection{Low-Complexity Algorithm}
The problem dimension of the proposed alternating algorithm could be prohibitive under wideband XL-MIMO. We therefore propose here a low-complexity sub-optimal solution, supported by the estimated angle-distance-delay parameters. Meanwhile, this scheme serves as a high-quality initial solution of the alternating algorithm. 

Specifically, unlike tackling sum rate maximization with full channel matrices, the principle of the low-complexity algorithm is to first design the analog beamformer to align with the estimated parameterized channel so that the channel gain is maximized, then fix the analog beamformer, and finally design the digital precoder to tackle the multi-user interference based on the low-dimensional effective channel.

Unlike most of the existing works that align the analog beamformer to the dominant path, we design the analog combiner of user $k$ to maximize the average combining gain of $L$ paths, which is favorable to FR3 frequency band.   We aim to solve  
\begin{align}
  \max _{\mathbf{f}_k} \; \sum\nolimits_{l=1}^L |\alpha_{k, l}|^2 \left|\mathbf{f}_k^ \herm \mathbf{b}\left(\psi_{k, l}\right)\right|^2 =  \mathbf{f}_k^ \herm  \mathbf{R}_{{\rm ue},k} \mathbf{f}_k ,
\end{align}
where $ \mathbf{R}_{{\rm ue},k} =  \sum_{l=1}^L |\alpha_{k, l}|^2  \mathbf{b}\left(\psi_{k, l}\right)  \mathbf{b}^\herm\left(\psi_{k, l}\right) $. As a Rayleigh quotient under constraint (\ref{ue_phase_constrinat}), the sub-optimal solution is 
\begin{align}\label{combiner_lc}
\mathbf{f}_k^\star =   \exp \left( j \arg \left( \mathbf{v}_{\max}(\mathbf{R}_{{\rm ue},k} ) \right)\right) /\sqrt{ {M_{\rm ue}}}.
\end{align}

Given $\mathbf{f}_k^\star$, we align the BS analog precoding to the average equivalent channels $\bar{\mathbf{h}}_k[n] = \mathbf{H}_k[n]\mathbf{f}_k^\star$ of $K$ users. The design criterion is therefore
\begin{align}\label{ave_gain}
\max _{\mathbf{W}_{\mathrm{RF}}} \;\sum_{k=1}^K \sum_{n=1}^N \omega_k \left\|\bar{\mathbf{h}}_k^\herm[n] \mathbf{W}_{\mathrm{RF}}\right\|^2
=  {\tr} (  
\mathbf{W}_{\mathrm{RF}}^\herm   \mathbf{R}_{{\rm bs}}  \mathbf{W}_{\mathrm{RF}}
),
\end{align}
where $\mathbf{R}_{\mathrm{bs}}=\sum_{k=1}^K \sum_{n=1}^N \omega_k\bar{\mathbf{h}}_k[n] \bar{\mathbf{h}}_k^\herm[n]$, and weight $\omega_k={1}/{ \sum_{n=1}^N\left\|\bar{\mathbf{h}}_k[n]\right\|^2}$ is used to make all user equal priority in calculating the average channel gain (\ref{ave_gain}).
Exploiting the block-diagonal structure of $\mathbf{W}_{\mathrm{RF}}$, we obtain the sub-optimal  solution of $\mathbf{W}_{\mathrm{RF}}^\star={\sf blkdiag}\left(\mathbf{w}_{\mathrm{RF}, 1}^\star, \ldots, \mathbf{w}_{\mathrm{RF}, M_{\mathrm{rf}}}^\star\right)$, where
\begin{align}
\mathbf{w}_{\mathrm{RF}, m_{\rm rf}}^\star= \exp \left(j \arg \left(\mathbf{v}_{\max }\left(\mathbf{R}_{\mathrm{bs}, m_{\rm rf}}\right)\right)\right) / \sqrt{\beta},
\end{align}
with $\mathbf{R}_{\mathrm{bs}, m_{\rm rf}} \triangleq \left[\mathbf{R}_{\mathrm{bs}}\right]_ {\mathcal{S}_{m_{\rm rf}}, \mathcal{S}_{m_{\rm rf}}} \in \mathbb{C}^{\beta \times \beta}$ and $\mathcal{S}_{m_{\rm rf}}$ denoting the index set of antennas connected to the $m_{\rm rf}$-th RF chain.

Given $\mathbf{f}_k^\star$ and $\mathbf{W}_{\mathrm{RF}}^\star$, the effective channel of user $k$ is 
\begin{align}
    \widetilde{\mathbf{H}}_k \triangleq  \left(  \mathbf{I}_N \otimes (\mathbf{f}_k^\star)^\herm\right)  \mathbf{H}_k^\herm  \left( \mathbf{I}_N \otimes \mathbf{W}_{\mathrm{RF}}^\star\right) \bC^{N\times M_{\rm rf}N}.
\end{align} 
Stacking all $K$ users' channel  as $\widetilde{\mathbf{H}} \triangleq [ \widetilde{\mathbf{H}}_1^\transp, \dots,  \widetilde{\mathbf{H}}_K^\transp]^\transp$,
a regularized zero-forcing (RZF) digital precoder is given as
\begin{align}
\mathbf W_{\rm BB}^\star =
\rho_{\rm zf}
\widetilde{\mathbf H}^{\herm}
\left(
\widetilde{\mathbf H}\widetilde{\mathbf H}^{\herm}
+
\mu_{\rm zf}\mathbf I_{KN}
\right)^{-1},
\end{align}
where $\mu_{\rm zf}=\ {KN\sigma_{\rm d}^2}/{P_{\rm d}}$ is the regularization parameter,
and $\rho_{\rm zf}$ is a normalization factor to satisfy power constraint (\ref{Wbb_power}).
 
The overall computational complexity mainly comes from the eigenvalue decomposition and RZF precoding, which is on the order of $\mathcal{O}\left(K M_{\mathrm{ue}}^3+M_{\mathrm{rf}} \beta^3+K^2 M_{\mathrm{rf}} N^3+K^3 N^3\right)$.

\begin{remark}
Conventional coordinated BF for multi-user MIMO \cite{7160780} requires either the availability of global CSI at the transmitter (like our alternating algorithm in Section \ref{alternating_section}), or repeated online BS-user iterations to construct the precoders and combiners, which, beyond computational complexity, also incurs substantial back-and-forth feedback and control signaling. By contrast, applying the proposed non-iterative algorithm, the BS only needs to inform a few estimated channel parameters $\{|\alpha_{k,l}|^2,\psi_{k,l}\}_{l\in[L]}$ to user $k$ (rather than the combiner itself). Then, each user simply configure its own combiner according to \eqref{combiner_lc}, and the BS can conduct the subsequent precoder design by itself. Our scheme therefore avoids iterative adaptation between the BS and users, reducing both computational complexity and signaling overhead.
    \hfill $\lozenge$
\end{remark}

\section{Simulation Results}\label{sec:simulation}
In the simulation, 
for channel estimation,
we consider the following system parameters and grid sizes: $M_{\rm bs} = 64\times32=2048$ with $G_{\theta^{\rm a}
} = M_{\rm h} = 64$ and $G_{\theta^{\rm e}} = M_{\rm v} = 32$, $M_{\rm rf} = 128$ with $\beta = 16$, $G_{\psi} = M_{\rm ue} = 8$, $G_{\tau} = N = 64$ with $\Delta f = 256$ kHz and the carrier frequency $f_c = 7$ GHz. 
The analog phase shifter $\{\wv_{m,t}\}$ and $\{\fv_t\}$ are generated with i.i.d. random phase. 
For DL data transmission, we evaluate sum spectral efficiency $\sum_k R_k/N$ with $K=8$ and $N=32$ by default. 
The channel parameters are set as: $L = 4$, $\theta^{\rm a}, \theta^{\rm e}, \psi \in [-60^\circ, 60^\circ]$,  $\tau \in [0, 200]$ ns, $\alpha_l\sim \mathcal{CN}(0, 1)$, $r \in [0.1, 0.5]R_{\rm Ray}$ with $R_{\rm Ray} \approx 105$ m as the Rayleigh distance, and $G_r = 6$ distance points uniformly distributed in the inverse distance domain, i.e., $1/r$.

\begin{figure}[t]
    \centering
    \includegraphics[width=0.8\linewidth]{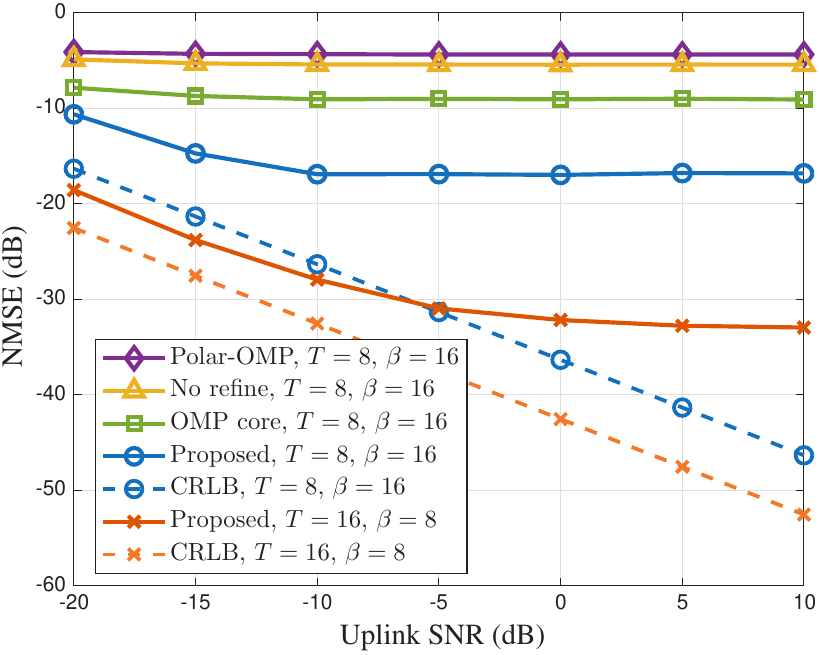}
    \caption{Channel estimation NMSE vs. UL SNR}
    \label{fig:mse}
\end{figure}

\begin{figure}[t]
    \centering
    \includegraphics[width=0.8\linewidth]{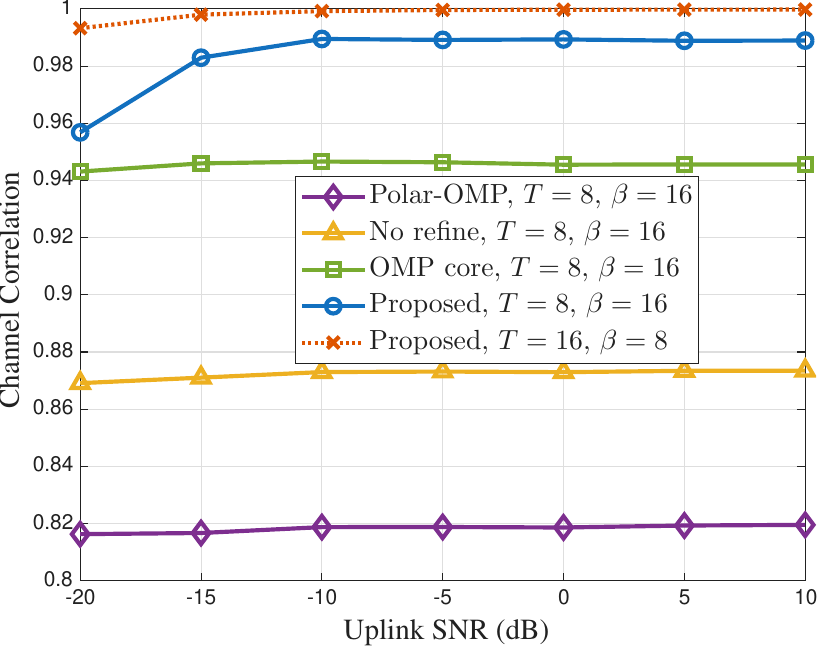}
    \caption{Channel estimation correlation vs. UL SNR}
    \label{fig:corr}
\end{figure}

\subsection{Channel Estimation Quality} 
Given $\hv$ and $\widehat{\hv}$ as vectorized true and estimated wideband channel, respectively, to evaluate the estimation performance, we consider NMSE $\bE[(\|\widehat{\hv}-\hv\|_2^2) / \|\hv\|_2^2]$ and channel correlation $\bE[|\widehat{\hv}^\herm \hv |/(\|\widehat{\hv} \|\|\hv\|)]$. 
We compare the proposed estimation algorithm with the following baselines:
\textbf{(1) OMP core:} This scheme replaces the SBL criterion $Q = |q|^2/s$ with the OMP matched-filter score $|q|^2$ throughout the algorithm, which corresponds to a decoupled off-grid Newtonized OMP (NOMP) and serves to demonstrate the advantage of the evidence-based SBL over OMP within the same decoupled framework. \textbf{(2) No refine:} This scheme is without the off-grid Newton refinement step, which quantifies the gain from continuous-domain parameter refinement and highlights the severity of grid mismatch. \textbf{(3) Polar-OMP:} This baseline uses a representative method based on the polar codebook of \cite{cui2022channel}, which operates on a joint angle-distance codebook and applies simultaneous OMP  across subcarriers. Since this scheme does not jointly estimate UE angle and delay within its codebook, we provide it with oracle information.  

The numerical results of NMSE and channel correlation are shown in Fig.~\ref{fig:mse} and Fig.~\ref{fig:corr}, respectively. 
We observe that the proposed full algorithm significantly outperforms the Polar-OMP baseline despite the oracle UE angle and delay information given in Polar-OMP. The ablation study shows that (i) the proposed continuous-domain Newton refinement is necessary, highlighted by the ``No refine" baseline; (ii) comparing with the ``OMP core" variant demonstrates the critical advantage of the SBL marginal likelihood criterion. Furthermore, by expanding the temporal observations ($T=16$) and increasing the spatial degrees of freedom with a denser RF configuration ($\beta=8$), the proposed algorithm acquires sufficient processing gain to effectively mitigate grid mismatch, tightly bounding the CRLB performance.

\subsection{Hybrid Precoding Quality} 
\begin{figure}[t]
    \centering
    \includegraphics[width=0.8\linewidth]{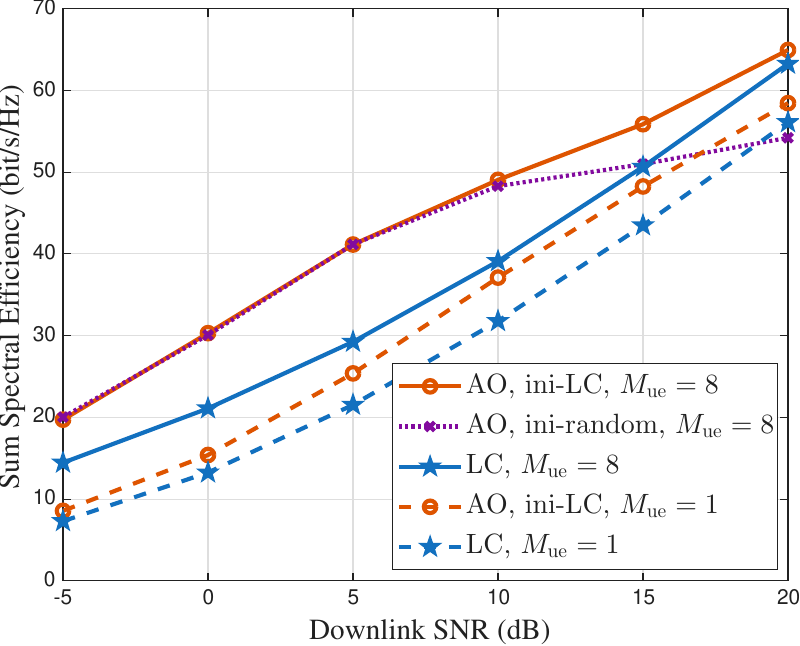}
    \caption{Achievable rate vs. DL SNR under $M_{\rm bs}=1024$, $\beta=8$.}
    \label{fig:downlinkrate1}
      \vspace{-5mm}
\end{figure}

\begin{figure}[t]
    \centering
    \includegraphics[width=0.9\linewidth]{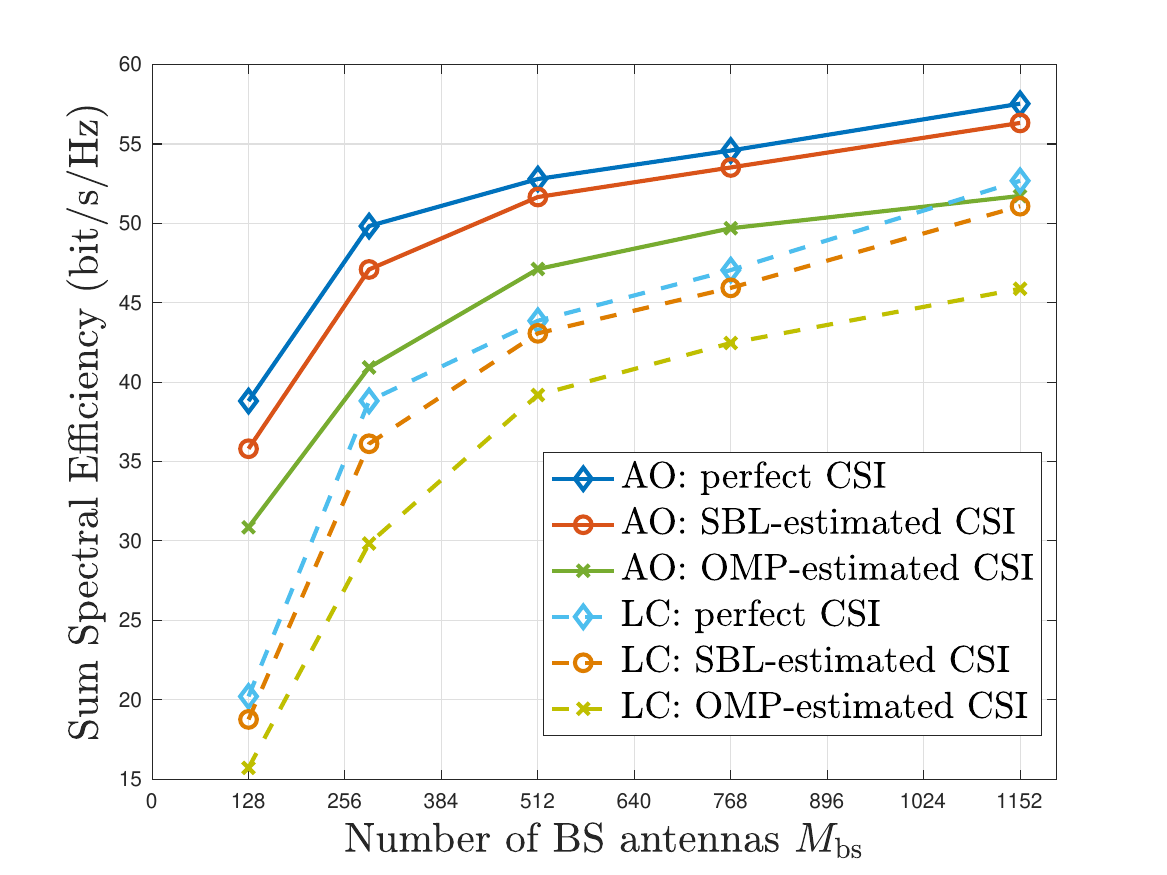}
    \caption{Achievable rate vs. $M_{\rm bs}$ with different CSI under 20 dB SNR.}
    \label{fig:downlinkrate2}
     \vspace{-5mm}
\end{figure}
Next, we examine the achievable rate performance achieved by the proposed alternating optimization (AO) and low-complexity (LC) optimization algorithms. We also consider AO initialized by LC (labeled as ``ini-LC'') and AO initialized randomly (labeled as ``ini-random''), respectively. 

The comparisons between AO and LC with perfect CSI under various DL SNRs are given in Fig.~\ref{fig:downlinkrate1}. It can be seen that both the proposed precoding algorithms can effectively enhance the communication performance given the near-field environments. Meanwhile, it can be seen that the proposed LC algorithm works well under the multipath condition (especially when $M_{\rm ue}$ is small) while it has much lower complexity compared to the AO algorithms. Besides, LC serves as a good initial solution for the AO algorithm when the solution space is prohibitively large and complicated. Considering the extremely large number of antennas for XL-MIMO operating in the multi-user and multi-subcarrier setup, limiting algorithm complexity is crucial. Therefore, the proposed LC algorithm holds great promise and practical feasibility for real-world systems involving a large number of optimization variables.

In Fig.~\ref{fig:downlinkrate2}, we evaluate the achievable rate performance based on estimated CSI. We observe that due to the high estimation accuracy of the proposed sequential SBL algorithm, the AO and LC algorithms based on the estimated CSI effectively approach the performance achieved by true CSI, even though the dimension of the channel is extremely large. This demonstrates the reliability of the proposed algorithm to realize transmission with an extremely high rate requirement. Besides, the advantages of the proposed algorithm compared to the OMP-core benchmark are showcased, where the gain comes from the more accurate parameter estimation, ensuring the accuracy of hybrid precoding.

\subsection{Subcarrier-Cooperative Gain} 
\begin{figure}
    \centering
    \includegraphics[width=1\linewidth]{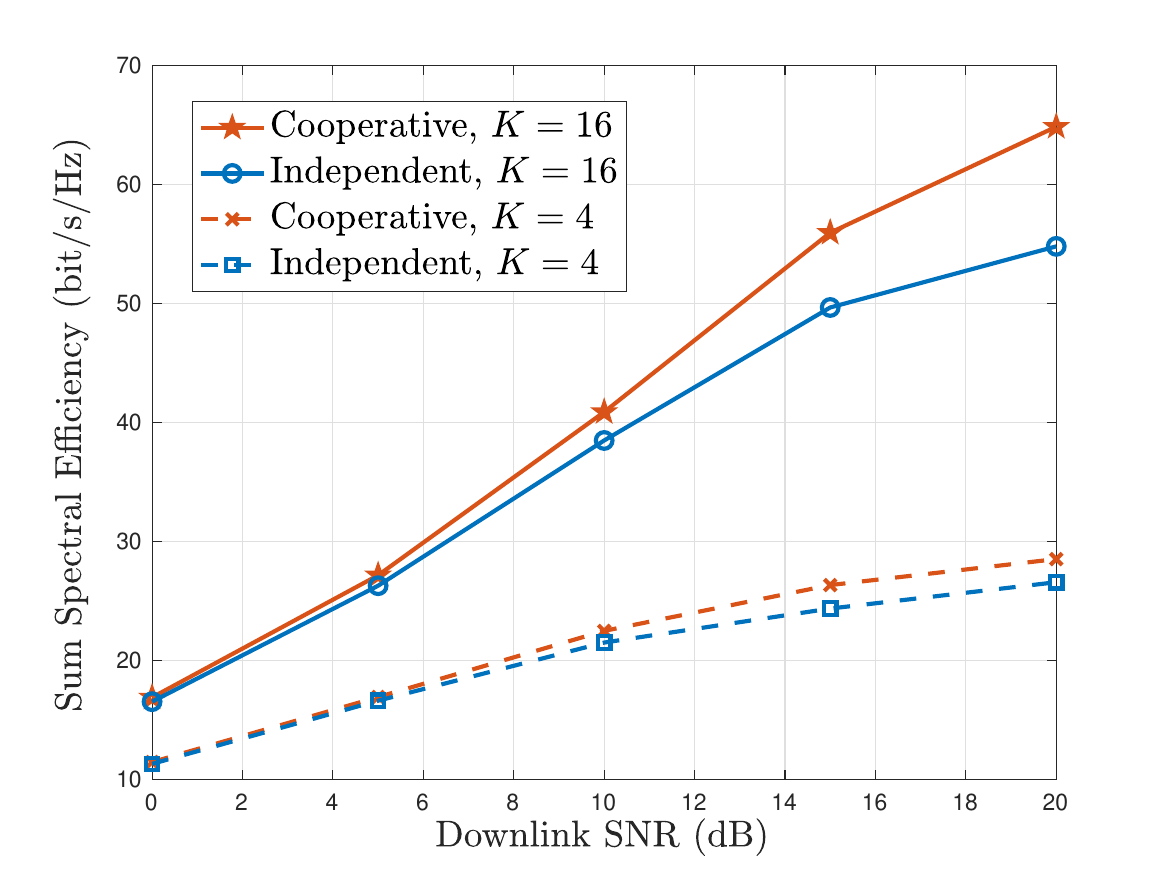}
    \caption{Achievable rate comparisons between wideband cooperative and independent BF schemes under $M_{\rm bs}=512$, $M_{\rm rf}=16$, $M_{\rm ue}=4$.}
    \label{fig:cooperGain}
      \vspace{-5mm}
\end{figure}
With perfect CSI and AO, Fig.~\ref{fig:cooperGain} illustrates the benefit of cooperative transmission over subcarriers. It is noticed that the gain of cooperative transmission mainly lies in the interference-limited regime, i.e., with high SNR and large $K$. This is because the wideband cooperation provides additional (approximately $\mathcal{O}(N)$) interference-nulling ability in the subcarrier domain. If strong multi-user interference exists, cooperative transmission has subcarrier-domain degrees of freedom to eliminate the interference and thus provide performance improvement. Otherwise, independent transmission \cite{yuan2023alternating} is sufficient to tackle the interference by spatial multiplexing.

\section{Conclusion}
In this paper, we proposed a decoupled off-grid sequential SBL algorithm to address the challenging near-field channel estimation problem in XL-MIMO OFDM systems.  
We effectively decomposed the computationally prohibitive 5-D joint parameter search into a sequence of low-dimensional marginal projections and proposed a continuous-domain Newton refinement to overcome severe grid mismatch. After acquiring CSI, two low-complexity hybrid-precoding algorithms were proposed to improve the transmission in the interference-limited regime with subcarrier cooperation. Simulations demonstrated the accuracy of estimation algorithms and communication performance of precoding based on estimated CSI in the challenging scenario of multi-user multi-subcarrier XL-MIMO.

\appendices
\section{Analytical First and Second Derivatives in \eqref{eq:newton_step}}\label{sec:newton_detail}

Let the first and second partial derivative of an atom $\phiv(\xiv)$ over each coordinate $\xi$ be $\phiv_\xi\triangleq \frac{\partial \phiv(\xiv)}{\partial \xi}$ and $\phiv_{\xi\xi}\triangleq \frac{\partial^2 \phiv(\xiv)}{\partial \xi^2}$ respectively, then we have $q_\xi  = \phiv_\xi^\herm \Cm^{-1}\yv$, $q_{\xi\xi}  = \phiv_{\xi\xi}^\herm \Cm^{-1}\yv$, $s_\xi = 2\Re\{\phiv_\xi^\herm \Cm^{-1}\phiv\}$ and $s_{\xi\xi} = 2\Re\{\phiv_{\xi\xi}^\herm \Cm^{-1}\phiv\} + 2\phiv_\xi^\herm \Cm^{-1}\phiv_\xi$. 
We further define $F(\xi)\triangleq |q(\xi)|^2$ and have $F_\xi = 2\Re\{q_\xi q^*\}$, $F_{\xi\xi} = 2|q_\xi|^2 + 2\Re\{q_{\xi\xi}q^*\}$.  
Then, $Q_\xi$ and $Q_{\xi\xi}$ are given as
\begin{align}    
    Q_\xi = \frac{F_\xi s -F s_\xi}{s^2},\quad 
    Q_{\xi\xi}  = \frac{F_{\xi\xi}}{s} -\frac{F s_{\xi\xi}}{s^2} - \frac{2F_\xi s_\xi}{s^2} + \frac{2F s_\xi^2}{s^3}. \notag
\end{align}
Now, we need to calculate $\phiv_{\xi}$ and $\phiv_{\xi\xi}$. Recall from \eqref{eq:atom3} that the atom entry before vectorization can be factored as $\mathcal{A}_{m,n,t}(\xiv) = g_n(\tau)v_{m,t}(\theta^{\rm a},\theta^{\rm e},r)u_t(\psi)$,
the partial derivative of $\phiv(\xiv)=\vec(\mathcal{A})$ is obtained by differentiating only the factor that depends on the selected coordinate.
\textbf{For delay coordinate} $\xi=\tau$: $\mathcal{A}'_{m,n,t}(\tau)=g'_n(\tau) v_{m,t}u_{t}$, $\mathcal{A}''_{m,n,t}(\tau)=g''_n(\tau) v_{m,t}u_{t}$, $g'_n(\tau) = -j2\pi(n-1)\Delta f g_n(\tau)$, $g''_n(\tau) = -(2\pi(n-1)\Delta f)^2 g_n(\tau)$.
\textbf{For UE angle coordinate} $\xi = \psi$: $\mathcal{A}'_{m,n,t}(\psi)=g_n v_{m,t}u'_{t}(\psi)$, $\mathcal{A}''_{m,n,t}(\psi)=g_n v_{m,t}u''_{t}(\psi)$, $u_t'(\psi) = \fv_t^\herm \bv_\psi$, $u_t''(\psi) = \fv_t^\herm \bv_{\psi\psi}$, $[\bv_\psi]_m = -j\pi(m-1)\cos(\psi)[\bv(\psi)]_m$, $[\bv_{\psi\psi}]_m = \big(j\pi(m-1)\sin(\psi) +(j\pi(m-1)\cos(\psi))^2\big)[\bv(\psi)]_m$.
\textbf{For one of BS angle and distance coordinates} $x\in\{\theta^{\rm a},\theta^{\rm e},r\}$: $\mathcal{A}'_{m,n,t}(x)=g_n v'_{m,t}(x)u_{t}$, $\mathcal{A}''_{m,n,t}(x)=g_n v''_{m,t}(x)u_{t}$, $v'_{m,t}(x) =[\Wm_t^\herm \av_x]_m$, $v''_{m,t}(x) =[\Wm_t^\herm \av_{xx}]_m$, $[\av_x]_m =[\av(x)]_m(-jk_0 d'_m(x))$, $[\av_{xx}]_m =[\av(x)]_m\left((-jk_0 d'_m(x))^2-jk_0d''_m(x)\right)$,
where $d_m(x) \triangleq \|\dv_m\triangleq \pv_m^{\rm bs}-\pv^{\rm sc}(x)\|$. Then, we further have 
\begin{align}
    d'_m(x) = -\frac{\dv_m^\transp \pv^{\rm sc}_x}{d_m(x)},\;
    d''_m(x)=\frac{\|\pv^{\rm sc}_x\|^2-\dv^\transp_m\pv^{\rm sc}_{xx}}{d_m(x)} - \frac{(\dv_m^\transp\pv^{\rm sc}_x)^2}{d^3_m(x)}, \notag
\end{align}
which are valid for any coordinate convention, as long as $\pv^{\rm sc}_x$ and $\pv^{\rm sc}_{xx}$ are computed consistently. For the case with UPA, we can write $\pv^{\rm sc}(\theta^{\rm a}, \theta^{\rm e}, r) = r \ev(\theta^{\rm a},\theta^{\rm e})$, where $\ev(\theta^{\rm a},\theta^{\rm e}) = [\cos(\theta^{\rm e})\cos(\theta^{\rm a}), \cos(\theta^{\rm e})\sin(\theta^{\rm a}), \sin(\theta^{\rm e})]^\transp$.
Then, $\ev'(\theta^{\rm a}) = [-\cos(\theta^{\rm e})\sin(\theta^{\rm a}), \cos(\theta^{\rm e})\cos(\theta^{\rm a}), 0]^\transp$, $\ev'(\theta^{\rm e}) = [-\sin(\theta^{\rm e})\cos(\theta^{\rm a}), -\sin(\theta^{\rm e})\sin(\theta^{\rm a}),\cos(\theta^{\rm e})]^\transp$, $\ev''(\theta^{\rm a}) = [-\cos(\theta^{\rm e})\cos(\theta^{\rm a}), -\cos(\theta^{\rm e})\sin(\theta^{\rm a}), 0]^\transp$ and $\ev''(\theta^{\rm e}) = -\ev$.
Thus, $\pv^{\rm sc}_{\theta^{\rm a}} = r \ev'(\theta^{\rm a})$, $\pv^{\rm sc}_{\theta^{\rm a}\theta^{\rm a}} = r \ev''(\theta^{\rm a})$, $\pv^{\rm sc}_{\theta^{\rm e}} = r \ev'(\theta^{\rm e})$ and $\pv^{\rm sc}_{\theta^{\rm e}\theta^{\rm e}} = r \ev''(\theta^{\rm e})$.  
For inverse distance $\rho = 1/r$, we have $\pv^{\rm sc}(\rho) = \rho^{-1}\ev$ and $\pv^{\rm sc}_{\rho} = -\rho^{-2}\ev$, $ \pv^{\rm sc}_{\rho\rho} = 2\rho^{-3}\ev$.

\section{Matrix-Inversion-Free Posterior Update}\label{sec:rank-one}
\textbf{Derivation of LOO posterior covariance $\Sigmam_{-i}$:}
We first define the posterior precision of active set $\Lambdam_{\bar{\bS}} \triangleq \Sigmam_{\bar{\bS}}^{-1} = N_0^{-1} \Phim^\herm_{\bar{\bS}}\Phim_{\bar{\bS}}+\Gammam^{-1}_{\bar{\bS}}$.  
Then for atom $i\in \bar{\bS}$,
we write the full posterior quantities in blocks as
\begin{align}
\Lambdam_{\bar{\bS}} = \begin{bmatrix}
        \Lambdam_{-i} &\lambdav_{\bar{i}} \\
        \lambdav_{\bar{i}}^\herm & \lambda_i
    \end{bmatrix},\;
    \Sigmam_{\bar{\bS}} = \begin{bmatrix}
        \Sigmam_{\bar{i}\bar{i}} &\sigmav_{\bar{i}} \\
        \sigmav_{\bar{i}}^\herm & \sigma_i
    \end{bmatrix}, \; \muv_{\bar{\bS}} = \begin{bmatrix}
        \muv_{\bar{i}} \\
        \mu_{i}
    \end{bmatrix},
\end{align}
where for notation simplicity, we define $\Lambdam_{\bar{i}\bar{i}} \triangleq [\Lambdam_{\bar{\bS}}]_{\bar{i},\bar{i}}$, $ \lambdav_{\bar{i}} \triangleq [\Lambdam_{\bar{\bS}}]_{\bar{i},i}$, $\lambda_i \triangleq [\Lambdam_{\bar{\bS}}]_{i,i}$,  $\Sigmam_{\bar{i}\bar{i}} \triangleq [\Sigmam_{\bar{\bS}}]_{\bar{i},\bar{i}}$, $ \sigmav_{\bar{i}} \triangleq [\Sigmam_{\bar{\bS}}]_{\bar{i},i}$, $\sigma_i \triangleq [\Sigmam_{\bar{\bS}}]_{i,i}$, $\muv_{\bar{i}} \triangleq [\muv_{\bar{\bS}}]_{\bar{i}}$, $\mu_i \triangleq [\muv_{\bar{\bS}}]_{i}$, and we notice that $ \Lambdam_{\bar{i}\bar{i}} = N_0^{-1}\Phim^\herm_{-i}\Phim_{-i} + \Gammam^{-1}_{-i} = \Lambdam_{-i}$.
We want to calculate $\Sigmam_{-i}$, which is the inversion of block $\Lambdam_{-i}$, i.e., $\Sigmam_{-i} = \Lambdam_{-i}^{-1}$. Using the block inverse formula of $\Lambdam_{\bar{\bS}}$ with $\lambda_{\rm sc}$ as the Schur complement of block $\Lambdam_{-i}$, we have 
\begin{align}
    \Lambdam^{-1}_{\bar{\bS}} = \begin{bmatrix}
        \Lambdam^{-1}_{-i}+ \Lambdam^{-1}_{-i} \lambdav_{\bar{i}} \lambda_{\rm sc}^{-1} \lambdav_{\bar{i}}^\herm \Lambdam^{-1}_{-i} & -\Lambdam^{-1}_{-i} \lambdav_{\bar{i}} \lambda^{-1}_{\rm sc}\\
        -\lambdav_{\bar{i}}^\herm \Lambdam^{-1}_{-i}  \lambda^{-1}_{\rm sc} & \lambda^{-1}_{\rm sc}
    \end{bmatrix},
\end{align}
in which each block equals to the corresponding block in $\Sigmam_{\bar{\bS}}$ due to  $\Lambdam_{\bar{\bS}}^{-1} = \Sigmam_{\bar{\bS}}$, i.e., 
\begin{subequations}\label{eq:block_equality}
\begin{align}
    \Sigmam_{\bar{i}\bar{i}}&=\Lambdam^{-1}_{-i}+ \Lambdam^{-1}_{-i} \lambdav_{\bar{i}} \lambda_{\rm sc}^{-1} \lambdav_{\bar{i}}^\herm \Lambdam^{-1}_{-i}, \label{eq:block1}\\
    \sigmav_{\bar{i}}&=-\Lambdam^{-1}_{-i} \lambdav_{\bar{i}} \lambda^{-1}_{\rm sc}, \quad \sigma_i =\lambda_{\rm sc}^{-1}. \label{eq:block3}
\end{align}
\end{subequations}
With $\Sigmam_{-i}=\Lambdam^{-1}_{-i}$, eliminating the cross terms in \eqref{eq:block_equality} yields 
\begin{align}
    \Sigmam_{-i} = \Sigmam_{\bar{i}\bar{i}} - \sigmav_{\bar{i}}\sigmav_{\bar{i}}^\herm/\sigma_i,
\end{align}
which proves the LOO posterior covariance in \eqref{eq:loo_sigma}.

\textbf{Derivation of LOO posterior mean $\muv_{-i}$:} First define $\gv_{\bar{\bS}}\triangleq N_0^{-1} \Phim_{\bar{\bS}}^\herm \yv$, then   $\muv_{\bar{\bS}} = \Sigmam_{\bar{\bS}} \gv_{\bar{\bS}}$ and $\gv_{\bar{\bS}} = \Lambdam_{\bar{\bS}} \muv_{\bar{\bS}}$, given in blocks 
\begin{align}\label{eq:g_block}
    \begin{bmatrix}
        \gv_{\bar{i}}\\ g_i
    \end{bmatrix} = \begin{bmatrix}
        \Lambdam_{-i} &\lambdav_{\bar{i}} \\
        \lambdav_{\bar{i}}^\herm & \lambda_i
    \end{bmatrix} 
    \begin{bmatrix}
        \muv_{\bar{i}} \\
        \mu_{i}
    \end{bmatrix},
\end{align}
where $\gv_{\bar{i}}\triangleq [\gv_{\bar{\bS}}]_{\bar{i}}$ and $g_i \triangleq [\gv_{\bar{\bS}}]_i$. Note that $\muv_{-i} = \Sigmam_{-i} \gv_{\bar{i}}$ and $\Sigmam_{-i} = \Lambdam^{-1}_{-i}$, with the first block row in \eqref{eq:g_block}, we have 
\begin{align}\label{eq:mu_middle}
    \muv_{-i} = \muv_{\bar{i}} + \Lambdam_{-i}^{-1} \lambdav_{\bar{i}} \mu_i.
\end{align}
From \eqref{eq:block3} we have $\Lambdam^{-1}_{-i}\lambdav_{\bar{i}} = -\sigmav_{\bar{i}} / \sigma_i$, which is combined with \eqref{eq:mu_middle} to yield
\begin{align}
    \muv_{-i} = \muv_{\bar{i}} - \sigmav_{\bar{i}}\mu_i/\sigma_i, 
\end{align}
which proves the LOO posterior mean in \eqref{eq:loo_mu}.

\textbf{Derivation of new posterior after adding new atom:}
The same block matrix inversion identity is used when adding a newly selected atom or adding back an updated atom after LOO refinement. To have a general derivation, we define $\bB$ as the base active set, where in forward addition we have $\bB = \bar{\bS}$, and in active-set refinement we have $\bB = \bar{\bS}_{-i}$. After adding an atom $\phiv_{\rm new} = \phiv(\xiv_{\rm new})$ with position $\xiv_{\rm new}$ and hyperparameter $\gamma_{\rm new}>0$, the new active set is $\bar{\bB} = \bB\cup\{(\xiv_{\rm new}, \gamma_{\rm new})\}$. Then, we define the augmented precision matrix in blocks as 
\begin{align}\label{eq:augment}
    \Lambdam_{\rm new} = \begin{bmatrix}
        \Lambdam_\bB & \cv\\
        \cv^\herm & c
    \end{bmatrix},
\end{align}
where $\cv \triangleq N_0^{-1}\Phim^\herm_\bB \phiv_{\rm new}$ and $c \triangleq \gamma_{\rm new}^{-1} + N_0^{-1}\|\phiv_{\rm new}\|^2$. We want to calculate $\Sigmam^{\rm new }_{\bar{\bB}} = \Lambdam_{\rm new}^{-1}$. Using the block inversion of $\Lambdam_{\rm new}$ with $\Sigmam_\bB = \Lambdam_\bB^{-1}$ and $\chi = c - \cv^\herm \Sigmam_\bB \cv$ as the Schur complement of $\Lambdam_\bB$, we have
\begin{align}
    \Sigmam^{\rm new}_{\bar{\bB}}= \Lambdam_{\rm new}^{-1} \!=\! \begin{bmatrix}
        \Sigmam_\bB + \Sigmam_\bB \cv \chi^{-1}\cv^\herm \Sigmam_\bB & - \Sigmam_\bB \cv \chi^{-1}\\
        -\cv^\herm \Sigmam_\bB \chi^{-1}& \chi^{-1}
    \end{bmatrix},
\end{align}
which proves the posterior covariance update for appending one atom in \eqref{eq:new_sigma}.

Next, the augmented posterior mean satisfies
\begin{align}\label{eq:augmented_mean}
\Lambdam_{\rm new} \muv^{\rm new}_{\bar{\bB}}=
 \begin{bmatrix}
        \Lambdam_{\bB} &\cv \\
        \cv^\herm & c
    \end{bmatrix} 
    \begin{bmatrix}
        \muv_\bB^{\rm new} \\
        \mu_{\rm new}
    \end{bmatrix}  =
    \begin{bmatrix}
        \gv_{\bar{\bB}}\\ g_{\rm new}
    \end{bmatrix},
\end{align}
where $g_{\rm new} = N_0^{-1} \phiv^\herm_{\rm new}\yv$. With $\muv_\bB = \Sigmam_\bB \gv_\bB$ and the first block row in \eqref{eq:augmented_mean}, we obtain 
\begin{align}\label{eq:mu_B_new}
    \muv^{\rm new}_\bB = \muv_\bB - \mu_{\rm new}\Sigmam_\bB \cv,
\end{align}
which is combined with the second block row in \eqref{eq:augmented_mean} to yield
\begin{align}\label{eq:mu_new}
    \mu_{\rm new} &= \chi^{-1}\left(g_{\rm new} - \cv^\herm \muv_\bB\right).
\end{align}
Putting \eqref{eq:mu_B_new} and \eqref{eq:mu_new} back to \eqref{eq:augmented_mean}, we obtain the full augmented posterior mean $\muv_{\bar{\bB}}^{\rm new}$, which proves the posterior mean update in \eqref{eq:new_mu}. 
We further notice that the numerator and denominator of $\mu_{\rm new}$ can be respectively  written as $g_{\rm new} - \cv^\herm \muv_{\bB} = N_0^{-1}\phiv_{\rm new}^\herm (\yv - \Phim_\bB \muv_\bB) = q_{-i}(\xiv_{\rm new})$ and  $ \chi = \gamma^{-1}_{\rm new} + s_{-i}(\xiv_{\rm new})$ under active-set refinement with $\bB = \bar{\bS}_{-i}$.  Hence, $\mu_{\rm new} = q_{-i}^{\rm new}/(\gamma^{-1}_{\rm new} + s_{-i}^{\rm new})$ proves \eqref{eq:mu_new_qs}.

{\small
	\bibliographystyle{IEEEtran}
	\bibliography{references}
}

\end{document}